\documentclass[preprint,aps,amssymb,showpacs,twocolumn,10pt,tightenlines]{revtex4}

\usepackage{amsmath,amssymb} 
\usepackage{psfig} 
\usepackage{epsfig}
\usepackage{psfrag}

\begin{document}
\DeclareGraphicsExtensions{.pdf,.png,.gif,.jpg}

\title{The Totally Asymmetric Simple Exclusion Process with Langmuir Kinetics}

\author{A. Parmeggiani$^{1,}$\footnote{New address: Laboratoire de
    Dynamique Mol\'eculaire des Interactions Membranaires, UMR 5539
    CNRS/Universit\'e de Montpellier 2, Place Eug\`ene Bataillon,
    34095 Montpellier Cedex 5, France.\\ email:
    parmeggiani@univ-montp2.fr, parmeggiani@hmi.de.}, T.
  Franosch$^{1,2}$, and E.  Frey$^{1,2}$}

\affiliation{$^1$ Hahn-Meitner Institut, Abteilung Theorie, Glienicker
  Str. 100, D-14109 Berlin, Germany\\ $^2$ Fachbereich Physik, Freie
  Universit\"at Berlin, Arnimallee 14, D-14195 Berlin, Germany}


\date{\today}

\begin{abstract}
  We discuss a new class of driven lattice gas obtained by coupling
  the one-dimensional totally asymmetric simple exclusion process to
  Langmuir kinetics.  In the limit where these dynamics are competing,
  the resulting non-conserved flow of particles on the lattice leads
  to stationary regimes for large but finite systems.  We observe
  unexpected properties such as localized boundaries (domain walls)
  that separate coexisting regions of low and high density of
  particles (phase coexistence). A rich phase diagram, with high an
  low density phases, two and three phase coexistence regions and a
  boundary independent ``Meissner'' phase is found.  We rationalize
  the average density and current profiles obtained from simulations
  within a mean-field approach in the continuum limit. The ensuing
  analytic solution is expressed in terms of Lambert $W$-functions. It
  allows to fully describe the phase diagram and extract unusual
  mean-field exponents that characterize critical properties of the
  domain wall.  Based on the same approach, we provide an explanation
  of the localization phenomenon. Finally, we elucidate phenomena that
  go beyond mean-field such as the scaling properties of the domain
  wall.
\end{abstract}

\maketitle

\section{Introduction}

\begin{figure}[htb]
\includegraphics[width=\columnwidth]{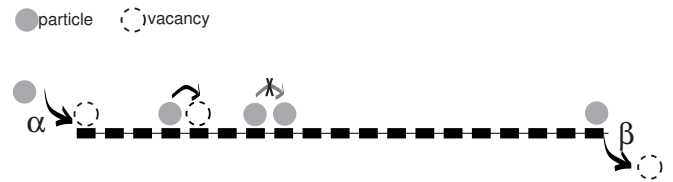}
\caption{Illustration of the Totally Asymmetric Simple Exclusion 
  Process with open boundaries. The entrance and exit rates at the
  left and right end of the one-dimensional lattice are given by
  $\alpha$ and $\beta$, respectively.}
\label{fig:tasep}
\end{figure}

Many natural phenomena driven by some external field or containing
self-propelled particles evolve into stationary states carrying a
steady current. Such states are characterized by a constant gain or
loss of energy, which distinguishes them from thermal equilibria.
Examples range from biological systems like ribosomes moving along
{\em m}-RNA or motor molecules ``walking'' along molecular tracks to
ions diffusing along narrow channels, or even cars proceeding on
highways. In order to elucidate the nature of such non-equilibrium
steady states a variety of driven lattice gas models have been
introduced and studied extensively~\cite{schmittmann_zia:95}. Here we
focus on one-dimensional (1D) models, where particles preferentially
move in one direction.  In this context, the Totally Asymmetric
Simple Exclusion Process (TASEP) has become one of the paradigms of
non-equilibrium physics~(for a review see Refs.~\cite{spohn:book,
  derrida_evans:review, mukamel:review, schuetz:review}).  In this
model a single species of particles is hopping unidirectionally and
with a uniform rate along a 1D lattice.  The only interaction between
the particles is hard-core repulsion, which prevents more than one
particle from occupying the same site on the lattice; see
Fig.~\ref{fig:tasep}.

It has been found that the nature of the non-equilibrium steady state
of the TASEP depends sensitively on the boundary conditions. For
periodic boundary conditions the system reaches a steady state of
constant density. Interestingly, density fluctuations are found to
spread faster than diffusively~\cite{beijeren_kutner_spohn:85}. This
can be understood by an exact mapping~\cite{meakin_etal:86} to a
growing interface model, whose dynamics in the continuum limit is
described in terms of the KPZ equation~\cite{kardar_parisi_zhang:86}
and its cousin, the noisy Burgers
equation~\cite{forster-nelson-stephen:77}.  In contrast to such ring
systems, open systems with particle reservoirs at the ends exhibit
phase transitions upon varying the boundary conditions~\cite{krug:91}.
This is genuinely different from thermal equilibrium systems where
boundary effects usually do not affect the bulk behavior and become
negligible if the system is large enough. In addition, general
theorems do not even allow equilibrium phase transitions in
one-dimensional systems at finite temperatures (if the interactions
are not too long-range)~\cite{landau_lifshitz_stat_mech:book}.

Yet another difference between equilibrium and non-equilibrium
processes can be clearly seen on the level of its dynamics. If
transition rates between microscopic configurations are obeying
detailed balance the system is guaranteed to evolve into thermal
equilibrium~\footnote{For stochastic processes described in terms of
  Langevin equations there is a set of potential
  conditions~\cite{deker-haake:75} which guarantees that the
  stationary state of the system is described by a Gibbs measure.}.
Systems lacking detailed balance may still reach a steady state, but
at present there are no universal concepts like the Boltzmann-Gibbs
ensemble theory for characterizing such non-equilibrium steady states.
In most instances one has to resort to solving nothing less than its
full dynamics. It is only recently, that exact (non-local) free energy
functionals for driven diffusive systems have been
derived~\cite{derrida_lebowitz_speer:01, derrida_lebowitz_speer:02}.

\begin{figure}[htb]
\includegraphics[width=\columnwidth]{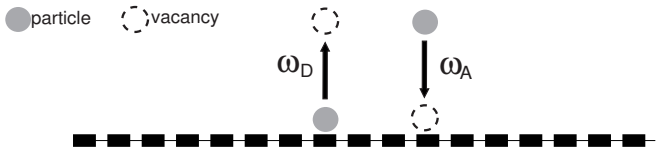}
\caption{Illustration of Langmuir kinetics. $\omega_A$ and $\omega_D$ 
  denote the local attachment and detachment rates.}
\label{fig:langmuir}
\end{figure}

This has to be contrasted with dynamic processes such as the
adsorption-desorption kinetics of particles on a lattice coupled to a
bulk reservoir (``Langmuir Kinetics'', LK), see
Fig.~\ref{fig:langmuir}.  Here, particles adsorb at an empty site or
desorb from an occupied one. Microscopic reversibility demands that
the corresponding kinetic rates obey detailed balance such that the
system evolves into an equilibrium steady state, which is well
described within standard concepts of equilibrium statistical
mechanics. If interactions between the particles other than the
hard-core repulsion are neglected, the equilibrium density is solely
determined by the ratio of the two kinetic
rates~\cite{fowler_stat_mech:book}, as given by the Gibbs ensemble.

The TASEP and LK can be considered as two of the simplest paradigms
which contrast equilibrium and non-equilibrium dynamics and stationary
states. Langmuir kinetics evolves into a steady state well described
in terms of standard concepts of equilibrium statistical mechanics.
Driven lattice gases such as the TASEP evolve into a stationary
non-equilibrium state carrying a finite conserved current.  Whereas
such non-equilibrium steady states are quite sensitive to changes in
the boundary conditions, equilibrium steady states are very robust to
such changes and dominated by the bulk dynamics. In the TASEP the
number of particles is conserved in the bulk of the one-dimensional
lattice. It is only through the particle reservoirs at the system
boundaries that particles can enter or leave the system. In LK
particle number is not conserved in the bulk.  Particles can enter or
leave the system at any site. Depending on whether we consider a
canonical or grand canonical ensemble the lattice is connected to a
finite or infinite particle reservoir.  Unlike the steady state of the
TASEP, the equilibrium steady state of LK does not have any spatial
correlations.

Combining both of these processes may at first sight seem a trivial
exercise since one might expect bulk effects to be predominant in the
thermodynamic limit.  This is indeed the case for attachment and
detachment rates, $\omega_A$ and $\omega_D$, which are independent of
system size $N$.  For large but finite systems interesting effects can
only be expected if the kinetics from the TASEP and LK compete. Then,
as we have shown recently~\cite{parmeggiani_franosch_frey:03}, novel
behavior different from both LK and the TASEP appears. In particular,
one observes {\em phase separation} into a high and low density domain
for an extended region in parameter space.

When should one expect competition between bulk dynamics (LK) and
boundary induced non-equilibrium effects (TASEP)? Let's consider the
following heuristic argument. A given particle will typically spend a
time $\tau \sim 1/\omega_D$ on the lattice before detaching. During
this ``residence'' time the number of sites $n$ explored by the
particle is of the order of $n \sim \tau$.  Hence, for fixed
$\omega_D$, the fraction $n/N \sim 1/ (\omega_D N) $ of sites visited
by a particle during its walk on the lattice would go to zero as $N
\to \infty$. Only if we introduce a ``total'' detachment rate by
$\Omega_D=N\omega_D$ and keep it constant instead of $\omega_D$ as
$N\to\infty$ will the particle travel a finite fraction of the total
lattice size. Similar arguments show that a vacancy visits an
extensive number of sites until it is filled by attachment of a
particle if $\omega_A$ scales to zero as $\omega_A = \Omega_A/N$ with
a fixed ``total'' attachment rate $\Omega_A$. In other words,
competition will be expected only if the particles live long enough
such that their internal dynamics or the external driving force
transports them a finite fraction along the lattice before detaching.
Then, particles spend enough time on the lattice to ``feel'' their
mutual interaction and, eventually, produce collective effects. In
summary, competition between bulk and boundary dynamics in large
systems ($N \gg 1$) is expected if the kinetic rates $\omega_A$ and
$\omega_D$ decrease with increasing system size $N$ such that the {\em
  total rates} $\Omega_A$ and $\Omega_D$ with
\begin{eqnarray}
\label{eq:redrates} 
\Omega_{A} 
= \omega_{A} N\, , & \quad & \quad \Omega_D = \omega_D N
\end{eqnarray}
are kept constant with $N$.  

The competition between boundary and bulk dynamics is a physical
process that has, to our knowledge, not yet been studied in the context of driven
diffusive systems.  In previous models emphasis was put on the
analysis of boundary induced phenomena in driven gases of mono- or
multi-species of particles~\cite{lahiri_ramaswamy:97,
  lahiri_barma_ramaswamy:00, evans_etal:98, mukamel:review,
  schuetz:review, schuetz:review2}, in presence of interactions (see
e.g.  Ref.~\cite{katz_lebowitz_spohn:83,katz_lebowitz_spohn:84}),
disorder~\cite{tripathy_barma:97,tripathy_barma:98} or local
inhomogeneities~\cite{kolomeisky:98, mirin_kolomeisky:03}, particles
with sizes larger than the lattice spacing~\cite{lakatos_chou:03,
  shaw_zia_lee:03}, lattices with different geometries (e.g.
multi-lanes lattice gases~\cite{popkov_peschel:01}), or systems in
presence of several conservation laws (for a review see e.g.
Ref.~\cite{schuetz:review2}).

In this manuscript we explore the consequences of particle
  exchange with a reservoir along the track (LK) on the stationary
  density and current profiles and the ensuing phase diagram of the
  TASEP. A short account of our ideas has been given
  recently~\cite{parmeggiani_franosch_frey:03}, where we have
  introduced the model and have shown how our Monte Carlo results can
  be rationalized on the basis of a mean-field theory, which we also
  solved analytically. The purpose of the present manuscript is to
  give a complete and comprehensive discussion of the topic. We will
  present results from Monte-Carlo results for the full parameter
  range of the model including the particular case where on- and
  off-rates equal each other, which were left out in our short
  contribution~\cite{parmeggiani_franosch_frey:03} due to the lack of
  space.  In addition, we will give the full reasoning for the
  derivation and analytical solutions of our mean-field theory. Here,
  additional insight is gained by identifying a branching point that
  explains all the features of the density profiles and phase diagram
  analytically. In particular, we show that a new critical point
  organizes the topology of the diagram and leads to unexpected 
  phenomena already briefly discussed in
  Ref.~\cite{parmeggiani_franosch_frey:03}. In a recent work, Evans {\em et al.}~\cite{evans_etal:03} have rephrased the  
   mean-field analysis first given in Ref.~\cite{parmeggiani_franosch_frey:03} and reproduced some of our results.
  The mean-field equations are, however, left in their implicit form
  and thus miss the interesting features we will obtain from the
  identification of a branching point.

These effects differ from those known in reference models of
equilibrium and non-equilibrium statistical mechanics like LK and the
TASEP.  Indeed, the coupling between the TASEP and LK, as is was
introduced above, produces new phenomena and extends the interest
toward systems which break conservation law in a non-trivial way. As
we shall see in the next Section, these features emerge already at
level of properties of the microscopic dynamics in configuration space
described by the master equation. 

Recently a variant of our model has been suggested by Popkov {\it
    et al.} ~\cite{popkov_etal:03}. Upon supplementing the
  Katz-Lebowitz-Spohn model by Langmuir kinetics and analyzing it
  within the mean-field approach similar to \cite{parmeggiani_franosch_frey:03},
  an even richer scenario for the stationary density profile is
  obtained that includes the emergence of localized downward domain
  walls and the appearance of several 'shocks' separating three
  distinct phases. It is also noted in Ref.~\cite{popkov_etal:03} that in general it may be important to replace the mean-field current by the exact current in the stationary state. 
 
In addition to its fundamental importance for non-equilibrium physics
in general, competition between bulk dynamics and boundary effects are
ubiquitous in nature, in particular biological phenomena. The TASEP
has actually been introduced in the biophysical literature as a model
mimicking the dynamics of ribosomes moving along a messenger RNA
chain~\cite{macdonald_gibbs_pipkin:68}; for generalizations of these
studies, see the recent work in Refs.~\cite{shaw_zia_lee:03,chou:03}.
Also some aspects of intracellular transport show close resemblance
to our model. For example, processive molecular motors advance along
cytoskeletal filaments while attachment and detachment of motors
between the cytoplasm and the filament occur~\cite{howard:book}.
Typically kinetic rates are such that these motors walk a finite
fraction along the molecular track before detaching. This falls well
into the regime where we expect novel stationary states.  Recently, it
has been shown that such dynamics can be relevant for modeling the
filopod growth in eukaryotic cells produced by motor proteins
interacting within actin filaments~\cite{kruse_sekimoto:02}. Finally,
our model could also be relevant for studies of surface-adsorption and
growth in presence of biased diffusion or for traffic models with bulk
on-off ramps~\cite{chowdury_etal:00}.

Since our paper contains a rather comprehensive discussion of the
  topic, we will give a detailed outline to provide the reader with
  some guidance through the analysis.  In Section \ref{s:model} we
  define the model by its dynamic rules and make a connection to its
  stochastic dynamics on a network. Though the relation between
  stochastic dynamics and networks is interesting to fully understand
  the peculiar features of the model introduced by the combination of
  conserved dynamics and on/off kinetics, it may be skipped for the
  first reading. We then present the problem in terms of a Fock space
  formulation and discuss the symmetries of the model, both key
  features for the subsequent formulation of the mean-field theory.
  In Section \ref{s:hydromfa} we briefly discuss some technical
  details of the Monte-Carlo simulation. Then follows a key section of
  the manuscript, a detailed development of the mean-field
  approximation and the resulting ``Burgers''-like equations in the
  continuum limit. Here we also discuss a series of features of these
  equations which will turn out to be crucial for the understanding of the
  ensuing density and current profiles.
  
  In Sect.  \ref{s:analytic} an analytic solution of the continuum
  equations is derived and compared to simulation results. We start
  the discussion for the special case that on- and off-rates are
  identical.  Though simpler to analyze, this case is somewhat
  artificial as it requires a fine-tuning of the on- and off-rates.
  Generically, one expects on- and off-rates to differ. Then the
  mathematical analysis becomes significantly more complex. We are
  still able to give an explicit analytical solution in terms of
  so-called Lambert functions, which allows us to identify a branching
  point that explains all the features of the density profiles and
  phase diagram analytically. In particular, we find a special
  point that organizes the topology of the diagram. In Sect. \ref{s:dw} we discuss the
  properties of the domain wall characterizing the phase coexistence
  upon changes of the model parameters. In
  particular, we show that in the vicinity of the special point mentioned above the
  domain wall exhibits non-analytic behavior similar to a critical
  point in continuous phase transitions.  We derive the critical
  exponents and the scaling related to the amplitude and position of
  the domain wall.  A conclusion, Sect.  \ref{s:conclusions},
  summarizes our results and provides additional arguments on the
  phenomenon of phase coexistence.  Last, we discuss some
  discrepancies between the mean-field approach and the simulation
  results and discuss a possible reconciliation.

\section{The Model}
\label{s:model}

In this Section we are going to describe the model in some detail. We
will also put it into the context of network theories. This will help
us to pinpoint the differences between the TASEP and LK dynamics and
show how a model combing both aspects will lead to novel phenomena.
Finally, we briefly review the key ideas of the Fock space formulation
of stochastic particle dynamics. In later chapters this formulation
will be used for an analytic discussion of the model.

\subsection{Definition of the dynamic rules}

\begin{figure}[htb]
\includegraphics[width=\columnwidth]{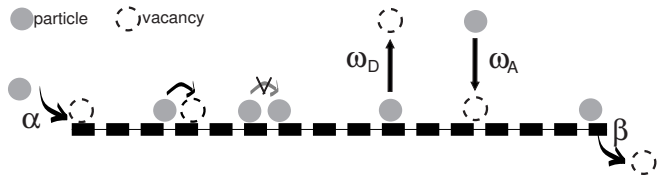} 
\caption{Schematic drawing of the totally asymmetric simple exclusion
  process with bulk attachment and
  detachment~\cite{parmeggiani_franosch_frey:03}. The entrance and
  exit rates at the left and right end of the one-dimensional lattice
  are given by $\alpha$ and $\beta$, respectively; $\omega_A$ and
  $\omega_D$ denote the local attachment and detachment rates.}
\label{f:model}
\end{figure}

In the microscopic model we consider a finite one-dimensional lattice
with sites labeled $i\!=\!1,...,N$ (see Fig.~\ref{f:model}) and
lattice spacing $a=L/N$, where $L$ is the total length of the lattice.
The site $i=1$ ($i=N$) defines the left (right) boundary, while the
collection $i\!=\!2,...,N-1$ is referred to as the bulk.

The microscopic state of the system is characterized by a distribution
of identical particles on the lattice, i.e. by configurations ${\cal
  C}=\{ n_{i=1,..N} \}$, where each of the occupation numbers $n_i$ is
equal either to zero (vacancy) or one (particle).  We impose a hard
core repulsion between the particles, which implies that a double or
higher occupancy of sites is forbidden in the model.  The full state
space then consists of $2^N$ configurations.

The statistical properties of the model are given in terms of the
probabilities ${\cal P} ({\cal C},t)$ to find a particular
configuration ${\cal C}=\{n_i\}$ at time $t$. We consider the
evolution of the probabilities ${\cal P}$ described by a master
equation:
\begin{equation}
\label{eq:master}
\frac{d {\cal P}({\cal C},t)}{dt}
= \sum_{ {\cal C'} \neq {\cal C}} 
  \Big[ {\cal W}_{{\cal C'}\to{\cal C}} {\cal P}({\cal C'},t) 
       - {\cal W}_{{\cal C}\to{\cal C'}} {\cal P}({\cal C},t)
  \Big] \, .
\end{equation}
Here, ${\cal W}_{{\cal C}\to{\cal C^\prime}}$ is a non-negative
transition rate from configuration ${\cal C}$ to ${\cal C^\prime}$.
As usual, master equations conserve probabilities.  The microscopic
processes connecting two subsequent configurations are local in
configuration space. Out of the possible $2^N \times 2^N$ transitions,
we consider only the following elementary steps connecting neighboring
configurations:
\begin{itemize}
\item[A)] at the site $i\!=\!2,..,N-1$ a particle can jump to site
  $i+1$ if unoccupied with unit rate;
\item[B)] at the site $i=1$ a particle can enter the lattice with rate
  $\alpha$ if unoccupied;
\item[C)] at the site $i=N$ a particle can leave the lattice with rate
  $\beta$ if occupied.
\end{itemize}
Additionally, in the bulk we assume that a particle: 
\begin{itemize} 
\item[D)] can leave the lattice with site-independent detachment rate
  $\omega_D$;
\item[E)] can fill the site (if empty) with a rate $\omega_A$ by
  attachment.
\end{itemize}
Processes $A)$ to $C)$ constitute a totally asymmetric simple
exclusion process with open
boundaries~\cite{derrida_evans:review, mukamel:review,
  schuetz:review}, while processes $D)$ and $E)$ define a Langmuir
kinetics~\cite{fowler_stat_mech:book}. We have taken the
attachment and detachment rates to be independent of the particle
concentration in the reservoir, i.e. we have assumed that the Langmuir
kinetics on the lattice is reaction and not diffusion limited. The
effect of diffusion in confined geometry has been studied in
Ref.~\cite{klumpp-lipowsky:03}.  A schematic graphical representation
of the resulting totally asymmetric exclusion model with Langmuir
kinetics~\cite{parmeggiani_franosch_frey:03} is given in
Fig.~\ref{f:model}.

Once we know the dynamic rules of the stochastic process, one may
introduce the notion of neighboring configurations for ${\cal C}$ and
${\cal C^\prime}$, if they differ only by a small fraction (${\cal
  O}(1/N)$) of the corresponding occupation numbers. This naturally
leads us to a reinterpretation of the dynamics in terms of networks as
described in the following subsection.

\subsection{Stochastic dynamics and networks}

The Markovian dynamics of the system can be represented in terms of a
network (graph), where the configurations of the stochastic process
correspond to the nodes (vertices) of the network. Each transition
allowed by the dynamics is represented as a directed link (edge), and
weighted by the corresponding transition rate which can be read off
from the dynamic rules $A)-E)$.  Due to the local dynamics the network
is very dilute.  A given node in the network is connected to a maximum
number $O(N)$ of nearby configurations.  Nevertheless, any
configuration can still be reached from any point within the network.
In other words the network is connected and does not break into
disjunct pieces. In addition, every node has at least one ingoing and
one outgoing link.  This guarantees that the system is ergodic, at
least as long as $N$ is finite, and all states are
recurrent~\cite{shiryaev:book}.

\begin{figure}[htbp!]
\includegraphics[width=\columnwidth]{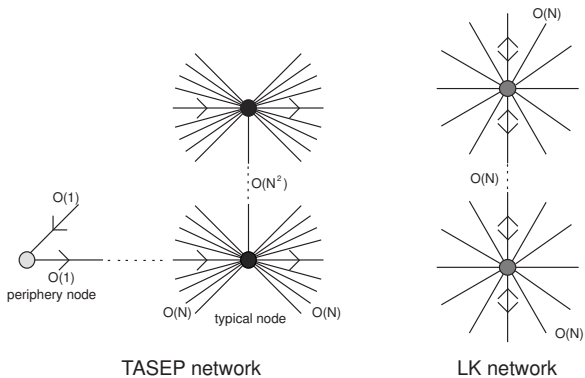}
\caption{Illustration of the network architecture corresponding to
  the totally asymmetric simple exclusion process (TASEP) and Langmuir
  kinetics (LK).}
  \label{f:topology}
\end{figure}

On such a network a distance between two different configurations can
be defined as the minimal number of steps required to connect them.
Note that the ``architecture'' of the network corresponding to a pure
TASEP is very different from a pure LK; see Fig.~\ref{f:topology} for
an illustration. 

The TASEP network is characterized by large fluctuations in the
connectivity. Take for example the completely filled configuration.
This state can only be left if the particle at the right end of the
lattice is ejected from the system. Similarly, a configuration
described by a step function $n_i = \Theta (x_i-x_0)$ with a
completely filled lattice to the left and a completely empty lattice
to the right of $x_0$ can only be left by a single process where the
rightmost particle is hopping forward. We call such and similar states
``periphery states'' since they are linked to the rest of the network
by a single or only a few outgoing and ingoing links. This is to be
contrasted with ``typical states'' for a given density, where
particles are more or less randomly distributed over the lattice.
Then, the conditional probability that an empty site is in front of a
filled site will be finite. In other words, there will be an extensive
number of pairs $(1,0)$ on the lattice. This implies that a typical
state will be connected with an extensive number $O(N)$ of directed
ingoing and outgoing links to other nodes in the network. Similarly,
the shortest path connecting two non-neighboring configurations has a
broad length distribution. Given two randomly chosen sequences of
occupation numbers $n_i=0$ and $n_i=1$ (i.e. nodes) one has to ask,
how many local moves of the type A) to C) (i.e. links) are needed to
transform one sequence into the other.  In general, there will be a
distribution of paths connecting these nodes. The shortest connection
may be only a few links, if local rearrangements of particles are
sufficient for matching the microscopic configurations. It seems
plausible that this is the case for such microscopic configurations,
whose coarse-grained density profiles are identical or at least very
similar. If the spatial profiles of the coarse-grained densities
corresponding to the two configurations differ significantly, one
expects $O(N^2)$ local rearrangements to be necessary for matching the
microscopic configurations. This is simply a consequence of particle
conservation in the bulk. For example, to completely empty a totally
filled state obviously requires $O(N^2)$ steps. In addition, distances
between two configurations in a TASEP network can also be highly
asymmetric.  Consider a configuration ${\cal C}$ corresponding to a
node at the periphery of the network connected to a configuration
${\cal C'}$.  Then the corresponding reverse step does not exist, and
in order return to the configuration ${\cal C}$ one has to make a
large loop in configuration space. In summary, a network corresponding
to TASEP contains only directed links. A characteristic feature is
its heterogeneity in the connectivity of nodes and distances between
nodes. The network contains loops, many of which may be very long due
to the conservation law in the bulk.

This has to be contrasted with the architecture of a network
corresponding to LK. Here, the connectivity of all nodes is
independent of the particular configuration. Since each occupation
number $n_i$ at a given site $i$ can be independently changed, the
number of links outgoing from a node is simply $N$.  To each outgoing
link there is an ingoing link with weights related by detailed
balance.  Moreover, any two configurations can be reached by at most
$N$ transitions. Since there is no conservation law, only local moves
(particle attachment or detachment) are necessary. The distance of two
configurations (along the shortest path) in a LK network is $d({\cal
  C},{\cal C'})=\sum_{i=2}^{N-1} |n_i - n_i'|$~\footnote{The left
  ($i=1$) and right ($i=N$) boundaries do not belong to the bulk where
  the Langmuir kinetics takes place.}. Since the order of the
necessary attachment and detachment processes is irrelevant the number
of such shortest paths is highly degenerate, and depends only on the
distance as $d!$. In summary, the LK network is not directed, very
homogeneous, highly connected and contains many loops of all size.

An important distinction between LK and the TASEP can be clearly seen
if one compares the nature of the corresponding stationary states.
Langmuir kinetics has a solution described in terms of the
thermodynamic equilibrium distribution:
\begin{equation} 
{\cal P}({\cal C}) = \frac{K^{|{\cal C}|}}{(K+1)^{N-2}} \, .
\end{equation}
Here $|{\cal C}| \equiv \sum_{i=2}^{N-1} n_i$ is the number of
occupied sites in the bulk and $K=\omega_A/\omega_D$ is the {\em
  binding constant} .  Note that the equilibrium distribution of LK
can be characterized by a Boltzmann weight upon introducing an
effective Hamiltonian ${\cal H}=-k_B T \sum_{i=2}^{N-1} n_i \ln K $.
The case $K=1$ has an interesting topological interpretation since the
links in the LK network loose their directionality and the effective
Hamiltonian ${\cal H}$ evaluates to $0$.

In contrast, the totally asymmetric exclusion process does not satisfy
the detailed balance condition
\begin{eqnarray*}
{\cal W}_{\,{\cal C^\prime}\to{\cal C}}~{\cal P}({\cal C^\prime}) =
{\cal W}_{\,{\cal C}\to{\cal C^\prime}}~{\cal P}({\cal C}) \, ,
\end{eqnarray*}
and evolves into a non-equilibrium steady state. Actually, if one
would assume detailed balance along a closed directed loop in the
TASEP network, one would be lead to the conclusion that all
probabilities along the path have to be zero. This, in turn, would
contradict the ergodicity of the finite system.

The network analogy discussed above can be used to understand why a
stochastic dynamics combining the totally asymmetric exclusion process
and Langmuir kinetics is interesting and show a range of novel
features not contained in the TASEP or LK alone. We have seen that the
number of links necessary to connect two non-neighboring states in the
TASEP ($O(N^2)$) is much larger than in LK ($O(N)$). Then, if we take
both the weights for hopping and the weights for attachment and
detachment to scale the same way, LK dynamics will dominate due to its
higher connectivity. In order to have competition, the weight of each
LK link has to be decreased as prescribed in the Introduction such
that the weighted path lengths of the TASEP and LK are comparable. Yet
another way to generate competition would be to only allow a finite
(non-extensive) number of sites to cause attachment and detachment
with a system size independent rate~\cite{kouyos_frey:unpub}. The
network structure of the totally asymmetric exclusion process with
Langmuir kinetics also indicates why standard matrix product ansatz
methods could be rather difficult to implement.

\subsection{Fock space formulation of stochastic dynamics}

It is sometimes convenient to formulate problems in stochastic
particle dynamics in terms of a {\em quantum Hamiltonian
  representation} instead of a master equation. This formalism was developed already some time ago by several groups~\cite{doi:76,
  grassberger_scheunert:80, peliti:84}. In the meantime it has found a
broad range of applications (see, e.g. Ref.~\cite{taeuber:03}).  We
refer the reader for details to various review
articles~\cite{schuetz:review,taeuber:03} and lecture
notes~\cite{cardy:notes,taeuber:notes}. \\
In our case, the occupation numbers $n_i({\cal C})$ constitute in a
natural way state space functions by measuring whether site $i$ is
occupied ($n_i=1$) or not ($n_i=0$) in configuration ${\cal C}$.  The
corresponding Heisenberg equations for ${\hat n}_i(t)$ then read
\begin{subequations}
\label{eq:micro}
\begin{eqnarray}
\label{eq:Dnew}
\frac{d}{dt} \hat{n}_i(t) 
& = & \hat{n}_{i-1}(t)\Big[ 1-\hat{n}_{i}(t)\Big] 
    - \hat{n}_i(t)\Big[1-\hat{n}_{i+1}(t)\Big] + \nonumber \\ 
& & +\, \omega_A\Big[1-\hat{n}_i(t)\Big] - \omega_D \hat{n}_i(t)
\end{eqnarray}
for any site in the bulk, while for sites at the boundaries one obtains
\begin{equation}
\label{eq:boundaries}
\begin{array}{l}
\displaystyle{\frac{d}{dt} \hat{n}_1(t) = 
\alpha\Big[1-\hat{n}_{1}(t)\Big] - \hat{n}_1(t) 
      \Big[1-\hat{n}_{2}(t)\Big]} \, ,\\
\\
\displaystyle{\frac{d}{dt} \hat{n}_N(t) = 
\hat{n}_{N-1}(t) \Big[1-\hat{n}_{N}(t)\Big] 
- \beta \hat{n}_N(t)}  \, .
\end{array}
\end{equation}
\end{subequations}
The first line of Eq.~(\ref{eq:Dnew}) is the usual contribution due to
the TASEP. Introducing the current operator
\begin{eqnarray*}
\hat{j}_i (t) = \hat{n}_i(t) \, \Big[1-\hat{n}_{i+1}(t)\Big] \, ,
\end{eqnarray*}
one can rewrite the right hand site of this line as $\hat{j}_{i-1} -
\hat{j}_i$, which is a discrete form of the divergence of the current.
This part defines a dynamics which satisfies particle number
conservation.  The second line of Eq.~(\ref{eq:Dnew}) represents the
additional Langmuir kinetics, which acts as source and sink terms in
the bulk.

These equations can now be understood as equations of motions for a
quantum many body problem. There are different routes to arrive at a
solution.  For one-dimensional problems there are many instances where
exact methods are applicable~\cite{schuetz:review}. Coherent state
path integrals are useful to explore the scaling behavior at critical
points~\cite{lee-cardy:95,cardy:notes,taeuber:03}. One can also try to
analyze the equations of motion
directly~\cite{mahan:book,gasser_elk:book}.  By taking averages of
Eqs. ~(\ref{eq:micro}) in order to compute the time evolution of
$\langle \hat{n}_i(t) \rangle$ one needs the corresponding averages of
two-point correlations such as $\langle \hat{n}_{i-1}(t)(1-\hat{n}_i(t))
\rangle$. This two-point correlation function obeys itself an equation
of motion connecting it to three-point and four-point correlation
functions. Thus we are lead to an infinite hierarchy of equations of
motion, as is quite generally the case for quantum many body
systems~\cite{gasser_elk:book,mahan:book}.  To proceed one can then
utilize standard approximation schemes of many body theory.

\subsection{Symmetries}

The system exhibits a particle-hole symmetry in the following sense. A
jump of a particle to the right corresponds to a vacancy move by one
step to the left.  Similarly, a particle entering the system at the
left boundary can be interpreted as a vacancy leaving the lattice, and
vice versa for the right boundary. Attachment and detachment of
particles in the bulk is mapped to detachment and attachment of
vacancies, respectively. Therefore, one can easily verify that the
transformation
\begin{subequations}
\begin{eqnarray}
\hat{n}_i(t) & \leftrightarrow & 1-\hat{n}_{N-i}(t) \\
\alpha & \leftrightarrow & \beta \\
\omega_A & \leftrightarrow & \omega_D
\end{eqnarray}
\end{subequations}
leaves Eqs. (\ref{eq:micro}) invariant. Due to this property we can
restrict the discussion to the cases $\omega_A > \omega_D $ and
$\omega_A=\omega_D$, i.e. to $K>1$ and $K=1$, respectively.
Eventually, for $\omega_A\!=\!\omega_D\!=\!0$, one arrives back at the
TASEP respecting the same particle-hole symmetry described above.

\section{Simulations, mean-field approximation and continuum limit}
\label{s:hydromfa}

In this Section we describe the Monte-Carlo simulations (MCS) and the
mean-field approximation (MFA) we have used to compute the stationary
average profile $\langle \hat{n}_i \rangle$ and the average current
$\langle \hat{j}_i \rangle = \langle \hat{n}_i (1 - \hat{n}_{i+1})
\rangle$.

\subsection{Simulations}

We have performed Monte-Carlo simulations with random sequential
updating using the dynamical rules $A)-E)$ and evaluated both time and
sample averages. The resulting profiles coincide in both averaging
procedures for given parameters and different system sizes. In the
simulations, stationary profiles have been obtained either over $10^5$
time averages (with a typical time interval $\ge 10~N$ between each
step of average) or over the same number of samples (in the case of
sample averages).

\subsection{Mean-field approximation and continuum limit}

Averaging Eqs. (\ref{eq:micro}) over the stationary ensemble relates
the mean occupation number to higher order correlation functions. The
mean-field approximation consists in neglecting these correlations
({\em random phase approximation}) \cite{gasser_elk:book,mahan:book}:
\begin{equation}
\langle \hat{n}_i(t) \hat{n}_{i+1}(t)\rangle 
= \langle \hat{n}_i(t) \rangle \; \langle \hat{n}_{i+1}(t) \rangle   \, .
\end{equation}
Here, averages in the stationary state $\langle \cdot \rangle$ are
actually time independent and correspond to either sample or time
averages due to the ergodicity property of the finite system.  In this
approximation the average current is given by 
\begin{eqnarray*}
\langle \hat{j}_i\rangle  = \langle \hat{n}_i(t) 
\rangle (1-\langle \hat{n}_{i+1}(t) \rangle) \, .  
\end{eqnarray*}
Once we have defined the average density at site $i$ as
$\rho_i = \langle \hat{n}_i(t) \rangle$, Eq.~(\ref{eq:Dnew}) results
in:
\begin{subequations}
\label{eq:micromf}
\begin{equation}
\label{eq:Dnewmf}
   \rho_{i-1}(1-\rho_{i}) - \rho_i(1-\rho_{i+1}) 
+  \omega_A(1-\rho_i) - \omega_D \rho_i = 0,
\end{equation}
while at the boundaries, Eqs. (\ref{eq:boundaries}), one obtains:
\begin{eqnarray}
\label{eq:boundariesmf}
\alpha(1-\rho_{1}) - \rho_1(1-\rho_{2}) =0\, , \nonumber \\
 \rho_{N-1} (1-\rho_{N}) - \beta \rho_N =0 \, .
\end{eqnarray}
\end{subequations}
Note that the average density is a real number with $ 0 \leq \rho_i
\leq 1$, and Eqs.~(\ref{eq:micromf}) form a set of $N$ real algebraic
non-linear relations, which can be solved numerically.

An explicit solution of the previous equations can be obtained by
coarse-graining the discrete lattice with lattice constant
$\varepsilon\! =\! L/N$ to a continuum, i.e. considering a {\em
  continuum limit}. To simplify notation, we fix the total length to
unity, $L\!=\!1$.  For large systems $N \gg 1$, $\varepsilon \ll 1$,
the rescaled position variable $x \equiv i/N$, $0 \leq x \leq 1$, is
quasi-continuous. An expansion of the average density $ \rho(x) \equiv
\rho_i$ in powers of $\varepsilon$ yields:
\begin{equation}
\rho(x \pm \varepsilon) 
= \rho(x) \pm \varepsilon\partial_x \rho(x) 
+ \frac{1}{2}\varepsilon^2\partial^2_x \rho(x)
+ O(\varepsilon^3) \, .
\end{equation}
Taking the scaling of the Langmuir rates, Eq.~(\ref{eq:redrates}),
into account, Eqs. (\ref{eq:micromf}) are to leading order in
$\varepsilon$ equivalent to the following non-linear differential
equation for the average profile at the stationary state
\cite{parmeggiani_franosch_frey:03}:
\begin{eqnarray}
\label{eq:mfa}
\frac{\varepsilon}{2}\partial_x^2 \rho 
+  (2 \rho -1)\partial_x \rho +  \Omega_A (1-\rho) 
-  \Omega_D \rho = 0 \, .
\end{eqnarray}
Equations (\ref{eq:boundariesmf}) now translate into boundary
conditions for the density field, $\rho(0)\!=\!\alpha$ and
$\rho(1)\!=\!  1-\beta$. This can be interpreted as if the system at
both ends is in contact with particle reservoirs of respective fixed
densities $\alpha$ and $1-\beta$. Note that the binding constant $K$
remains unchanged in this limit.

For finite $ \varepsilon$, the average current is written $j= -
\frac{\varepsilon}{2} \partial_x \rho + \rho(1-\rho)$. In the
continuum limit $\varepsilon \to 0^+$, this suggests that $j=\rho
(1-\rho)$ and that the current is bounded, $j \le 1/4$. However, this
bound holds only if the density is a smooth function of the position
$x$.  We shall show that density discontinuities can arise in the
continuum limit.  Then, for small $\varepsilon$, these discontinuities
would appear as rapid crossover regions where one cannot neglect the
first order derivative term in the current definition so that the
relation $j \le 1/4$ needs not to be satisfied.  The inequality can be
violated also by the additional contribution arising from current
fluctuations neglected in the mean-field approximation; see e.g. Fig.~
\ref{f:er6} at the system boundaries.
 
The equations obtained in mean-field approximation and the subsequent
continuum limit still respect the particle-hole symmetry mentioned
above. In terms of the continuous averaged density $\rho$, the symmetry now
reads $\rho(x) \mapsto 1 -\rho(1-x)$, $\alpha \!\leftrightarrow
\!\beta, \, \,\Omega_A \!\leftrightarrow \!\Omega_D$. Note that a
numerical solution of the differential equation above necessarily uses
a discretization. Using a standard algorithm for integrating
differential equations, one would merely recover the original
mean-field equations (\ref{eq:micromf}).

Equation (\ref{eq:mfa}) has mathematical similarities to the
stationary case of a viscous Burgers equation~\cite{burgers:39,
  zwillinger:book, zachmanoglou_thoe:book}
\begin{eqnarray}
\label{eq:mfagen} 
\partial_t \rho -  \frac{\varepsilon}{2} \partial_x^2 \rho 
+ (\partial_\rho j) \partial_x \rho
= {\cal F}_A - {\cal F}_D \, .
\end{eqnarray}
In the Burgers equation $\rho$ is identified with the fluid velocity and
$j$ is related to this velocity via $j=\rho^2/2$.  In our case, the
hard-core interaction between particles implies a non-linear
current-density relationship. As shown above, one finds in the
continuum limit a parabolic relation $j=\rho(1-\rho)$.  Dissipation is
due to the term $\varepsilon\partial_x^2 \rho$, while the sources
represent fluxes from and to the bulk reservoir ${\cal F}_A = \Omega_A
(1-\rho)$ and ${\cal F}_D = \Omega_D \rho$.  The net source term
${\cal F}_A - {\cal F}_D=(K+1)\Omega_D(\rho_l-\rho)$ is positive or
negative depending on whether the density $\rho$ is below or above the
Langmuir isotherm, $\rho_l=K/(K+1)$, expressed in terms of the binding
constant $K=\Omega_A/\Omega_D$.  In conjunction with the non-linear
current-density relation this implies that the density of the Langmuir
isotherm will act like an ``attractor'' or ``repellor''.  If the slope
of the current-density relation is positive, $\partial_\rho j >0$, and
the density at the left end falls below the Langmuir isotherm the bulk
reservoir will feed particles into the system. As result, the density
grows towards $\rho_l$ as one moves away from the boundary. In
contrast, for a negative slope $\partial_\rho j < 0$, i.e. for
densities larger than $1/2$, the density profiles are ``repelled''
from the Langmuir isotherm. The latter case can also be understood as
an ``attraction'' by the Langmuir isotherm if read starting from the
right end of the system.  Then, depending on whether the density at
the right boundary is larger or smaller than $\rho_l$ there is a loss
or gain of particles from the reservoir as one moves away from the
right boundary into the bulk.  This will turn out to be an important
principle for the discussion of the density profiles in later
sections; see e.g.  Section~\ref{sec:generic_case}.

{F}rom the analogy to fluid dynamics
problems~\cite{landau_lifshitz_fluid_mech:book} one expects
singularities such as shocks in the density $\rho$ to appear in the
inviscid or non-dissipative limit $\varepsilon \to 0^+$. This conclusion
can also be inferred by a direct inspection of the non-linear
differential equation (\ref{eq:mfa}) in the limit $\varepsilon=0$. It
reduces to a first order differential equation,
\begin{equation}
\label{eq:mfa0} 
(2 \rho -1)\partial_x \rho + 
\Omega_A (1-\rho) -\Omega_D \rho=0 \, ,
\end{equation} 
instead of a second order one, while the solution still has to satisfy
two boundary conditions.  Such a boundary value problem is apparently
{\em over-determined}. However, we can define solutions of
Eq.~(\ref{eq:mfa0}) respecting only one of the boundary conditions.
Depending on whether they obey the boundary conditions on the left or
right end of the lattice we call them the {\em left solution}
$\rho_\alpha$ and the {\em right solution} $\rho_\beta$, respectively.
Then, for $0 < \varepsilon \ll 1$ the full solution of Eq.~(\ref{eq:mfa}) 
will be close to $\rho_\alpha$ for positions on the
left side of the system and similarly to $\rho_\beta$ on the right
side.  In general, we can not expect both solutions to match
continuously at some point in the bulk of the lattice. Instead, for a
large but finite system, the solution of Eq.~(\ref{eq:mfa}) will
exhibit a rapid crossover from the left to the right solution.  In the
limit $\varepsilon \to 0^+$ this crossover regime decreases in width
and eventually leads to a discontinuity of the average density profile
at some position $x_w$. Note that the discontinuity shows up only on
the scale of the system size, i.e.  in the rescaled variable $x$,
whereas on the scale of the lattice spacing the crossover region
always covers a large number of lattice sites.

To locate the position of the discontinuity $x_w$ in the limit of
large system sizes $N \gg 1$, i.e. $\varepsilon \to 0^+$, it is useful
to derive a continuity equation for the current $j$ and the sources
${\cal F}_A,\,{\cal F}_D$. Consider Eq.~(\ref{eq:mfa}) in the form
$\partial_x j ={\cal F}_A - {\cal F}_D$, where $j = -
\frac{\varepsilon}{2} \partial_x \rho + \rho(1-\rho)$. Integrating
over a small region of width $2 \delta x$ close to $x_w$, one obtains
$ j(x_w + \delta x) - j(x_w - \delta x)= \int_{x_w - \delta x}^{x_w +
  \delta x}({\cal F}_A - {\cal F}_D) \, dx \equiv {\cal
  S}_\varepsilon$. In the limit $\varepsilon \to 0^+$ the relation
simplifies to $j_\alpha(x_w + \delta x) - j_\beta(x_w - \delta x) =
{\cal S}_0$, where we have defined the left current $j_\alpha =
\rho_\alpha (1-\rho_\alpha)$ and similarly for the right current
$j_\beta$. Now, for $\delta x \to 0^+$, the contribution due to the
sources ${\cal S}_0$ is of order $\delta x$ yielding the {\em matching
  condition} in terms of the left and right currents
\begin{equation}
\label{eq:continuity} j_\alpha(x_w) = j_\beta(x_w) \, .
\end{equation}
The equivalent condition for the densities reads
\begin{equation}
\label{eq:dwp1} \rho_\alpha(x_w)=1-\rho_\beta(x_w) \, .
\end{equation}
A discontinuity of the density profile such as a {\em domain wall} can
appear in the system depending on whether the previous condition is
fulfilled for $0 \le x_w \le 1$. Relation (\ref{eq:dwp1}), therefore,
defines implicitly where a domain wall is located in the system. It
allows to compute the domain wall position $x_w$ as well as its height
$\Delta_w=\rho_\beta(x_w)-\rho_\alpha(x_w)$. The domain wall separates
regions of low ($\rho < 1/2$) and high density ($\rho > 1/2$). In the
ensuing phase diagram this will lead to an extended regime of phase
coexistence.

We shall see that in addition to domain walls, there may appear also
discontinuities in the current~\footnote{Since we agreed that, for
  finite $\epsilon$, the current is defined as
  $j=-\frac{\varepsilon}{2} \partial_x \rho + \rho(1-\rho)$, we have to
  stress that the boundary layer becomes a discontinuity in the
  density and the current {\em in the continuum limit}, i.e. for
  $\varepsilon \to 0^+$.}, which are located at the boundary of the
system. We refer to them as {\em boundary layers}.

\section{Analytic solution of the continuum equation}
\label{s:analytic}

In this Section, we will show in detail how one can treat the
continuum equations, Eq.~(\ref{eq:mfa}), analytically in the limit
$\varepsilon \to 0^+$. We shall compare these results with numerical
solutions of Eq.~(\ref{eq:mfa}) for finite $\varepsilon$~\footnote{The
  numerical solution of the mean-field equation for finite
  $\varepsilon > 0$, Eq.~(\ref{eq:mfa}), including the boundary
  conditions have been obtained using standard routines available in
  Maple, release 7.} , and with corresponding profiles obtained from
Monte-Carlo simulations.  For the Monte-Carlo simulation the plots
will show the average density $\langle \hat{n}_i \rangle $ and the
average current $\langle \hat{j}_i \rangle = \langle \hat{n}_i
(1-\hat{n}_{i+1})\rangle$. The densities and currents obtained from
the numerical integration of the mean-field equations at finite
$\varepsilon$ will be indicated as $\rho_\varepsilon$ and
$j_\varepsilon = -\varepsilon/2 \partial_x \rho_\varepsilon +
\rho_\varepsilon (1-\rho_\varepsilon)$ in the figures, respectively.

This discussion will result in a classification of the possible
solutions as a function of the entry and exit rates $\alpha$ and
$\beta$, the binding constant $K=\Omega_A/\Omega_D$ and the detachment
rate $\Omega_D$ (phase diagram). Due to the particle-hole symmetry we
can restrict ourselves to values $K \geq 1$. Then, there are two cases
to distinguish: $K=1$ and $K > 1$. For $K=1$ the constant density
profile, $\rho_l=K/(K+1)$, given by the Langmuir kinetics coincides
with a point of particular symmetry of the TASEP.  Indeed, for a
density of $\rho = 1/2$ the system is dual under particle-hole
exchange, the non-linear term in Eq.~(\ref{eq:mfa}) vanishes, and it
corresponds to a point of maximal current~\footnote{Note that one can
  write for the non-linear term in Eq.~(\ref{eq:mfa}) $\partial_x j =
  \partial_x \rho \partial_\rho j = (2 \rho - 1) \partial_x \rho$,
  which implies $\partial_\rho j =0$ at density $\rho=1/2$.}. It will
turn out that $K=1$ introduces particular features and requires a
specific treatment and discussion.  Since it is technically simpler we
discuss this case first.

\subsection{The symmetric case: $K=1$}

The mathematical analysis is simplified by the fact that the
attachment and detachment rates are equal, $\Omega_A = \Omega_D \equiv
\Omega$. Then Eq.~(\ref{eq:mfa0}) factorizes to:
\begin{equation}
\label{eq:mfaer}
(2 \rho -1)(\partial_x \rho - \Omega)=0 \, .
\end{equation}
The boundary conditions read $\rho(0) = \alpha$ and $\rho(1) =
1-\beta$.  Note that this equation is symmetric with respect to
particle-hole exchange.  Indeed, except for the boundaries, the
equation is invariant under the transformation $\rho(x) \mapsto 1 -
\rho(1 - x)$. This has important consequences for the density
profiles, as will become clear in the following.

\subsubsection{The density and current profiles}

Equation (\ref{eq:mfaer}) has only two basic solutions. A constant
density $\rho_l (x) = 1/2$ identical to the stoichiometry in Langmuir
kinetics and also the density in the maximal current phase of the
TASEP.  The other solution is a linear profile $\rho = \Omega x + C$.
The value of the integration constant $C$ depends on the boundary
condition. One finds $C_\alpha = \alpha$ and $C_\beta = 1- \beta -
\Omega$ for solutions, $\rho_\alpha (x)$ and $\rho_\beta (x)$,
matching the density at the left and the right boundary, respectively.
Depending on how the three solutions $\rho_\alpha(x)$, $\rho_\beta(x)$
and $\rho_l (x)$ can be matched, different scenarios arise for the
full density profile $\rho(x)$. In the following we discuss the
characteristic features of the solution in each quadrant of the
$\alpha$--$\beta$ phase diagram for fixed $\Omega$.

\paragraph{Lower left quadrant: $\alpha, \beta \le 1/2$.} 
In this case the boundary conditions enforce a density less than $1/2$
and greater that $1/2$ at the left and right boundaries, respectively.
This allows for a continuous density profile, where a constant density
of $\rho_l=1/2$ intervenes between the two linear solutions emerging
from the left and right boundaries. The corresponding positions
separating the low density from the maximal current phase,
$\rho_\alpha(x_\alpha)=1/2$, and the maximal current phase from the
high density phase, $\rho_\beta(x_\beta)=1/2$, are given by $x_\alpha
= (1-2 \alpha)/2\Omega > 0$ and $x_\beta = (2\beta +2 \Omega
-1)/2\Omega<1$, respectively. The phase boundary $x_\alpha \to 0^+$
moves to the left upon increasing the entry rate $\alpha \to {1/2}^-$
and similarly $x_\beta \to 1^-$ for the exit-rate $\beta \to {1/2}^-$.
Hence, depending on the values of the points $x_\alpha$ and $x_\beta$,
one can classify the possible solutions according to the relative
ordering of the phase boundaries: (i) $x_\alpha < x_\beta$, (ii)
$x_\alpha = x_\beta $, and (iii) $x_\alpha > x_\beta$.

\begin{figure}[htbp!]
\includegraphics[width=0.7\columnwidth]{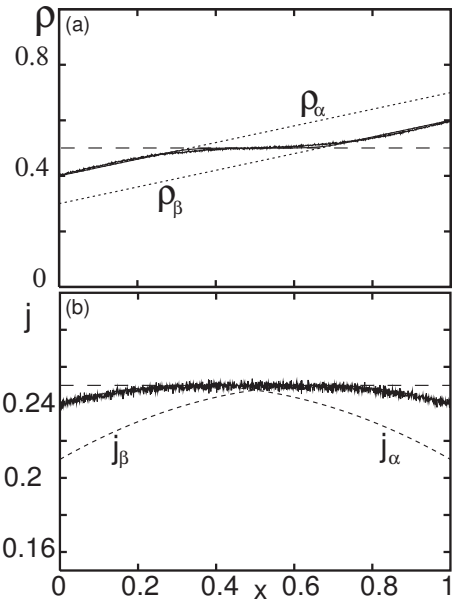}
\caption{Average density $\rho(x)$ {\bf (a)} and current $j(x)$ {\bf (b)}
  for parameters $\alpha=0.4$, $\beta=0.4$, $\Omega=0.3$ and $K=1$. In
  this parameter range one observes a 3-phase coexistence: a maximal
  current phase is intervening between a low and high density phase.
  The profiles are computed analytically in the inviscid limit (dashed
  lines) and numerically for $\varepsilon=10^{-3}$ within a mean-field
  approximation (solid smooth line), and from Monte-Carlo simulations
  (solid wiggly line). Note that, within the resolution of the
  figures, the Monte-Carlo results and the numerical mean-field
  results can not be distinguished. The analytic density profile is
  shown for the solutions respecting the left and right boundaries
  conditions, $\rho_\alpha$ and $\rho_\beta$; we also show the
  Langmuir isotherm $\rho_l=1/2$.}
\label{f:er1}
\end{figure}

(i) The density profile is continuous and piecewise linear and given by
\begin{equation}
\rho(x)= 
  \left\{ \begin{array}{lcr}
          \Omega x + \alpha & \text{for} & 0\le x \le x_\alpha \, ,\\
          1/2 &  \text{for}  & x_\alpha \le x \le x_\beta \, ,\\
          \Omega (x-1) + 1 - \beta &  \text{for}&  x_\beta \le x
          \le 1 \, .
         \end{array}
  \right.
\end{equation}
One observes a region of {\em 3}-phase coexistence: a low density
phase (LD) with $\rho(x)< 1/2$ and $j(x) < 1/4$ for $0\le x \le
x_\alpha$, a maximal current phase (MC) with $\rho(x) = 1/2$ and $j(x)
= 1/4$ for $x_\alpha \le x \le x_\beta$ and high density phase (HD)
with $\rho(x) > 1/2$ and $j(x) < 1/4$ for $x_\beta \le x \le 1$. For a
plot of the densities and currents see Fig.~\ref{f:er1}.

(ii) For $x_\alpha = x_\beta$ the width of the intermediate maximal
current phase vanishes and the solution becomes a simple linear
profile, continuously matching the densities of the LD and HD phase.

\begin{figure}[htbp!]
\includegraphics[width=0.7\columnwidth]{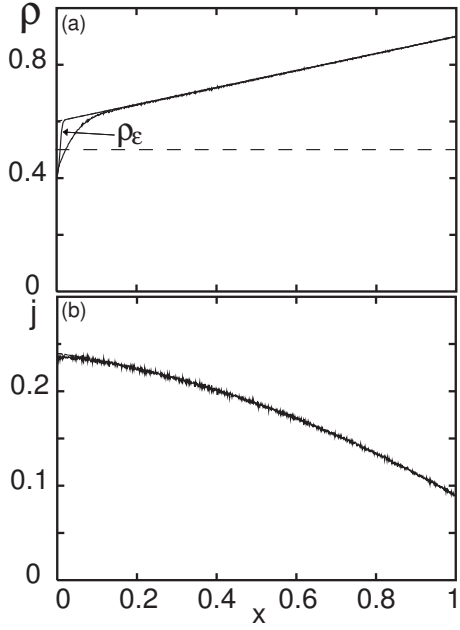}
\caption{Average density $\rho (x)$ {\bf (a)} and current $j (x)$ {\bf (b)}
  for parameters $\alpha=0.4$, $\beta=0.1$, $\Omega=0.3$ and $K=1$. We
  use the same legend as in Fig.~\ref{f:er1}.  The bulk profile is
  almost completely described by the solution $\rho_\beta$ matching
  only the right boundary condition. At the left end, the bulk density
  does not match the boundary condition.  As a result, a boundary
  layer appears. Only there does one find a noticeable difference
  between the profiles of the Monte-Carlo simulation, the numerical
  computation at finite $\varepsilon$ and the analytic profile for
  vanishing $\varepsilon$.}
\label{f:er2}
\end{figure}

(iii) Upon further increasing $x_\alpha$ over $x_\beta$, the
intervening maximal current phase is lost and it is no longer possible
to continuously concatenate the linear density profiles of the low and
high density phase. There is necessarily a density discontinuity,
located at a point $x_w$ where the currents corresponding to the right
and left solutions match, $j_\alpha(x_w) = j_\beta(x_w)$.  The
position of the ensuing domain wall may be in or outside of the
system. This leads us to further distinguish between the following
three subcases:

\indent\indent ($\mbox{iii}_{1}$) If $x_w<0$ the density profile in
the bulk is above $1/2$, i.e. in a HD phase. The profile is entirely
described by the solution $\rho_\beta(x)$ up to a boundary layer at
the left end. One observes that the boundary layer corresponds to a
discontinuity in the current. The bulk current $j_\beta (x \to 0^+)$
does in general not match the incoming particle flux $\alpha
(1-\alpha)$ at the left boundary (see Fig.~\ref{f:er2}).

\indent\indent($\mbox{iii}_{2}$) For $0<x_w<1$ the domain wall is
within the system boundaries. Then the density profile connects a LD
profile to a HD profile via a domain wall at position $x_w=(\Omega -
\alpha + \beta) / 2\Omega$~\footnote{In terms of the rate constants
  the condition $0<x_w<1$ reads $\alpha \le \beta + \Omega$ for $0 <
  x_w$ and $\alpha \ge \beta - \Omega$ for $x_w < 1$.}.  The density
profile is given by:
\begin{equation}
\rho(x)= \left\{
\begin{array}{lcr}
\Omega x + \alpha & \text{for} & 0\le x \le x_w \, ,\\
\Omega (x-1) + 1 - \beta &  \text{for} &  x_w \le x \le 1 \, .
\end{array}
\right.
\end{equation}

\begin{figure}[htbp!]
  \includegraphics[width=0.7\columnwidth]{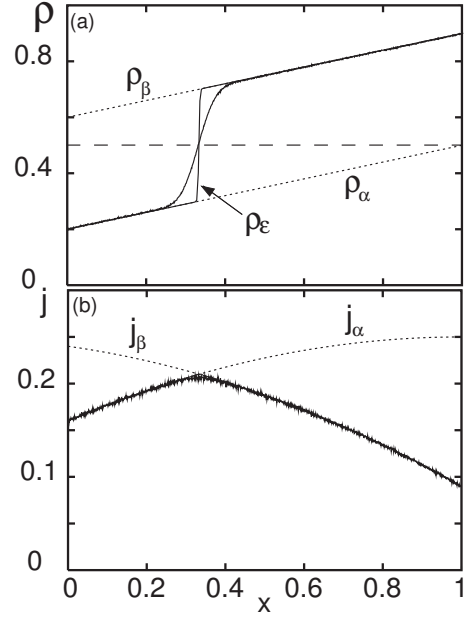}
\caption{Average density $\rho (x)$ {\bf (a)} and current $j (x)$ {\bf (b)}
   for parameters $\alpha=0.2$, $\beta=0.1$, $\Omega=0.3$ and
  $K=1$. We use the same legend as in Fig.~\ref{f:er1}.  Only in
  proximity to the domain wall the results from the mean-field
  approximation show deviations from the density profile obtained by
  Monte-Carlo simulation.}
\label{f:er3}
\end{figure}

Here we can already illustrate an important feature of our model.  As
one can infer from Fig.~\ref{f:er3}, the current forms a cusp at the
position of the domain wall, with $j_\alpha (x)$ and $j_\beta (x)$
being monotonically increasing and decreasing functions of $x$,
respectively.
This follows directly from the continuum equation,
Eq.~(\ref{eq:mfagen}), and the density dependence of the source term
${\cal F}_A - {\cal F}_D = 2 \Omega (1/2 - \rho) $, which is positive
or negative depending on whether the density is smaller or larger than
$1/2$. Hence, the domain wall is located at a maximum of the current.
In addition, the strict monotonicity of the current also implies that
the domain wall is {\em localized}. A displacement of the domain wall
to the right of $x_w$ would result in a current $j_\alpha > j_\beta$.
This in turn would increase the influx of particles at the left
boundary, which will drive the domain wall back to its original
position $x_w$ \footnote{See also the Section \ref{s:conclusions}.}.

\indent\indent($\mbox{iii}_{3}$) The solution for $x_w>1$ can be
inferred by particle-hole symmetry from case ($\mbox{iii}_{1}$). 
The low density profile is given by the solution $\rho_\alpha(x)$ 
up to a boundary layer at the right end. 

\begin{figure}[htbp!]
\includegraphics[width=\columnwidth]{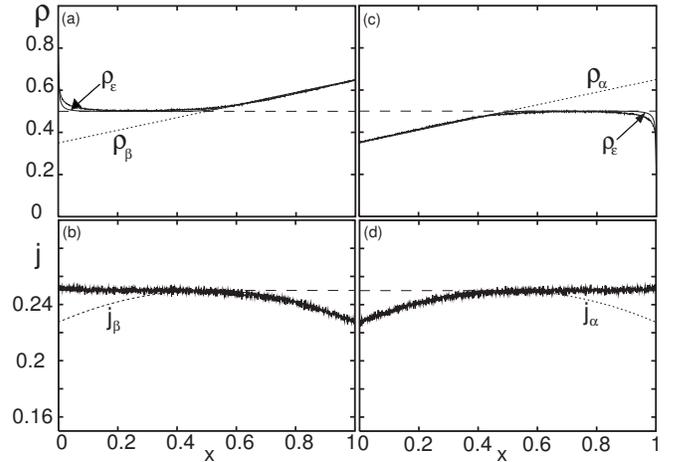}
\caption{{\bf (a)}-{\bf (b)} Average density $\rho (x)$ and current $j(x)$ 
  for $\alpha=0.8$, $\beta=0.35$, $\Omega=0.3$ and $K=1$.  
  We use the same legend as in Fig.~\ref{f:er1}. Except for the left boundary layer, in fig. {\bf (a)}
  the analytic solution is described by the Langmuir density
  $\rho_l=1/2$ and the density $\rho_\beta$ matching the right
  boundary condition. {\bf (c)}-{\bf (d)} Average density $\rho (x)$ and current $j(x)$ for $\alpha=0.35$, $\beta=0.8$  	and the same $\Omega$ and $K$ as before. Note that the curves map to those of {\bf (a)}-{\bf (b)} by
  particle-hole symmetry.}
\label{f:er4}
\end{figure}

\paragraph{Lower right quadrant: $\alpha > 1/2$, $\beta < 1/2$.}
Here the density at both left and right boundaries is larger
than $\rho_l = 1/2$. Two different scenarios are possible. In the
first scenario, the slope $\Omega$ of the density profile $\rho_\beta
(x)$ (matching the density at the right boundary) is so small that
$\rho_\beta (x)$ is always larger than $\rho_l = 1/2$; this requires
$\Omega < 1/2 -\beta$. Then, the bulk of the system is in the HD phase
with a boundary layer on the left. This scenario is identical to the
previous case ($\mbox{iii}_{1}$), such that there is no qualitative
change in the bulk upon crossing the line $\alpha=1/2$. In other
words, there is no phase boundary and the system remains in the HD
phase. In the second scenario, the slope $\Omega > 1/2 - \beta$ such
that we have a phase boundary between a high density and a maximal
current phase. This solution can also be viewed as a limit of the
3-phase coexistence region, where for $\alpha \to 1/2^-$ the phase
boundary $x_\alpha$ leaves the system through the left end and a
boundary layer is created replacing the LD region (see
Figs.~\ref{f:er4}(a,b)).

\paragraph{Upper left quadrant: $\alpha < 1/2$, $\beta > 1/2$.}
This region in parameter space is obtained using particle-hole
symmetry from the results for the lower right quadrant in the
preceding paragraph (see Figs.~\ref{f:er4}(c,d)).

\begin{figure}[htbp!]
\includegraphics[width=0.7\columnwidth]{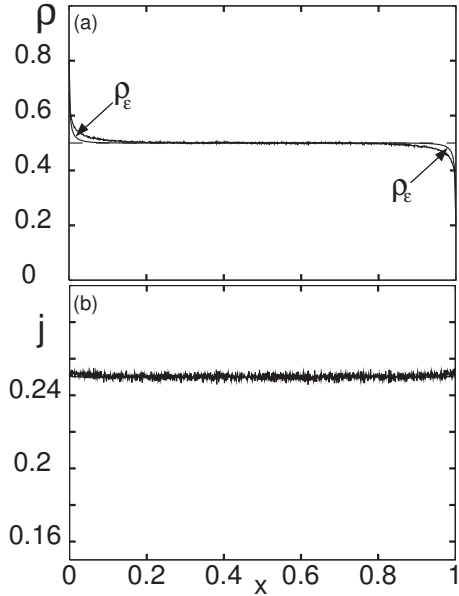}
\caption{Average density $\rho (x)$ {\bf (a)} and current $j (x)$ {\bf (b)}
  for parameters $\alpha=0.8$, $\beta=0.8$, $\Omega=0.3$ and $K=1$. We
  use the same legend as in Fig.~\ref{f:er1}.  The bulk density
  profile is given by the Langmuir density $\rho_l=1/2$ which
  corresponds also to the maximal current phase. Due to fluctuations,
  neglected in the mean-field approximation, the current profile
  obtained from the simulation exceeds the value $1/4$ at each
  boundary.}
\label{f:er6}
\end{figure}

\paragraph{Upper right quadrant: $\alpha, \beta > 1/2$.}
Here two boundary layers are formed, and the bulk of the system is
characterized by a constant density equal to $1/2$ (see
Fig.~\ref{f:er6}).  This corresponds to the maximal current phase,
which remains unchanged as compared to the TASEP without particle on-
and off-kinetics. Note again that due to $K=1$ the density with
maximal current coincides with the Langmuir isotherm $\rho_l =1/2$.

\subsubsection{The phase diagram}

\begin{figure}[htbp!]
\includegraphics[width=\columnwidth]{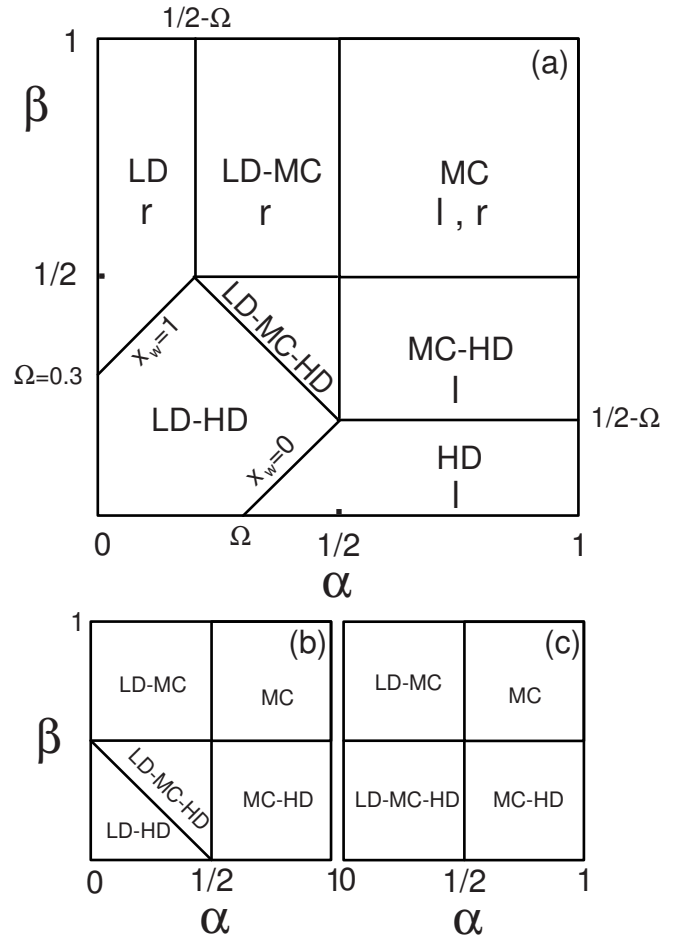}
\caption{Cut of the phase diagram on the  $(\alpha,\beta)$--plane 
  in the mean-field approximation for $K=1$ and different values of
  $\Omega$: {\bf (a)} $\Omega=0.3$, {\bf (b)} $\Omega=0.5$, {\bf (c)}
  $\Omega=1.0$. The cases (a)-(c) correspond to the three different
  topologies of phase diagrams discussed in the main text. All lines
  represent continuous transitions between different regions in the
  $(\alpha, \beta, K = 1, \Omega_D = \text{const.})$ cut of the
  $4$-dimensional parameter space.  The line parallel to the
  anti-diagonal is defined through the relation $\alpha + \beta
  +\Omega = 1$. It represents the border line where the points
  $x_\alpha$ and $x_\beta$ (i.e. the points where the left and right
  solutions $\rho_\alpha$ and $\rho_\beta$ meet the Langmuir isotherm
  $\rho_l=1/2$) coincide, $x_\alpha=x_\beta$.  The phase boundaries of
  the LD-HD coexistence phase, $x_w=0$ and $x_w=1$, correspond to
  regions in which the domain wall is located at one of the system
  boundaries.  These lines were computed by using the matching
  conditions for the currents: $j_\alpha(1) = \beta(1-\beta)$ and
  $j_\beta(0) = \alpha(1-\alpha)$.  In figure (a), we also emphasize
  the presence of the boundary layers at the left or the right end of
  the system. These are indicated with the letters "l" and "r",
  respectively. Such boundary layers remain present in the same
  regions as in figure (a) also for increasing $\Omega$.}
\label{f:erdiag1}
\end{figure}

The analysis of the current and density profiles allows to draw cuts
of the phase diagram in the $(\alpha,\beta)$--plane for fixed values
of $\Omega$. Note that the particle-hole symmetry renders all diagrams
symmetric with respect to the diagonal $\alpha=\beta$.  Depending on
the kinetic rate $\Omega$ one can distinguish three topologies.
Topologies of the phase diagrams change at critical values
$\Omega=1/2$ and $\Omega = 1$; see Fig.~\ref{f:erdiag1}.

For $0 <\Omega < 1/2$, Fig.~\ref{f:erdiag1}(a), the phase diagram
consists of seven phases. A 3-phase coexistence region LD-MC-HD at the
center is surrounded by three 2-phase coexistence regions LD-HD, MC-HD
and LD-MC.  Pure LD, HD and MC phases are contiguous to the 2-phase
regions. All lines between different regions represent continuous
changes in the average density $\rho$. The 3-phase coexistence region,
and two of the 2-phase coexistence regions (LD-MC and MC-HD) are
characterized by continuous density profiles. This is mainly due to
the maximal current phase with density $\rho_l (x) = 1/2$. Acting as a
"buffer", this phase intervenes between the LD and HD phase or
connects the LD and HD phases with the right and left boundary,
respectively. Discontinuities only appear as current and density
discontinuities (boundary layers) at the system boundaries.  This has
to be contrasted with the density profile in the coexistence region
between the LD and HD phase. Here, a density discontinuity in the bulk
(domain wall), separating both phases, is formed.

It is also interesting to consider the limit $\Omega \rightarrow 0$,
as one expects to recover the TASEP scenario.  Indeed, using the
previous results, it is easy to show that for decreasing $\Omega$, the
width of the 2-phase regions, as well as that of the 3-phase region, shrinks to
zero. The resulting diagram reproduces the well known topology of the
pure TASEP in the mean-field approximation
\cite{derrida_domany_mukamel:92}.

Upon increasing $\Omega$ up to the value $1/2$, we find the first
topology change in the phase diagram. The HD and LD phases gradually
disappear, leaving only the 2-phase or 3-phase coexistence regions at
$\Omega=1/2$, see Fig.~\ref{f:erdiag1}(b). If $\Omega$ becomes larger
than $1$, the LD-HD coexistence region disappears; see
Fig.~\ref{f:erdiag1}(c).

The Langmuir kinetics is approached for $\Omega \to \infty$. Although
the topology of the phase diagram does not change anymore, the phases
become almost indistinguishable for large kinetic rates. Here the
Langmuir isotherm $\rho_l = 1/2$ occupies most of the bulk, whereas
the LD and HD regions are confined to a vicinity of the boundaries.

\subsection{The generic case: $K>1$}
\label{sec:generic_case}

Though simpler to analyze, the previous case $K=1$ is somewhat
artificial as it requires a fine-tuning of the on- and off-rates.
Generally, one would expect $K \neq 1$. Due to particle-hole
symmetry we can restrict ourselves to $K > 1$. The analysis becomes
significantly more complex since the continuum equation for the
density, Eq.~(\ref{eq:mfa0}), no longer factorizes into a simple form
as for $K=1$.

\subsubsection{The density and current profiles}

To proceed, it is convenient to introduce a rescaled density of the
form
\begin{equation}
\label{eq:changecoord} 
\sigma(x) = \frac{K+1}{K-1} \left[ 2\rho(x)-1 \right] -1 \, , 
\end{equation}
where $\sigma = 0$ corresponds to the Langmuir isotherm
$\rho_l=K/(K+1)$.  Since the density $\rho (x)$ is bound within the
interval $[0,1]$, the rescaled density $\sigma (x)$ can assume values
within the interval $[-2K/(K-1),2/(K-1)]$. Then the continuum equation
Eq.~(\ref{eq:mfa0}) simplifies to
\begin{equation}
\label{eq:Bmod2} 
\partial_x\sigma(x) + \partial_x\ln|\sigma(x)|
= \Omega_D \frac{(K+1)^2}{K-1} \, .
\end{equation}
Direct integrations yield
\begin{equation}
\label{eq:lambrel} 
|\sigma(x)| \exp(\sigma(x)) = Y(x) \, ,  
\end{equation}
where the function $Y(x)$ is
\begin{equation}
Y(x) = |\sigma(x_0)| \exp\left\{\Omega_D\frac{(K+1)^2}{K-1} (x-x_0) 
+ \sigma(x_0) \right\} \, ,
\end{equation}
and $\sigma(x_0)$ is the value of the reduced density at the reference
point $x_0$.  In particular, the ones that match the boundary
condition on the left or right end of the system are written
\begin{eqnarray}
\label{eq:gencoord}
Y_\alpha(x) 
& = & | \sigma(0) |\exp\left\{\Omega_D\frac{(K+1)^2}{K-1} x 
+ \sigma(0) \right\} \\
\nonumber \\
Y_\beta(x) 
& = & \left|\sigma(1)\right|\exp\left\{\Omega_D \frac{(K+1)^2}{K-1}
  (x-1) + \sigma(1)\right\} \, , \nonumber  
\end{eqnarray}
where the boundary values $\sigma(0)$ and $\sigma(1)$ can be written
in terms of $\alpha$ and $\beta$ using Eq.~(\ref{eq:changecoord}) and
the boundary conditions $\rho(0)=\alpha$ and $\rho(1)=1-\beta$.
 
\begin{figure}[htb]
  \includegraphics[width=\columnwidth]{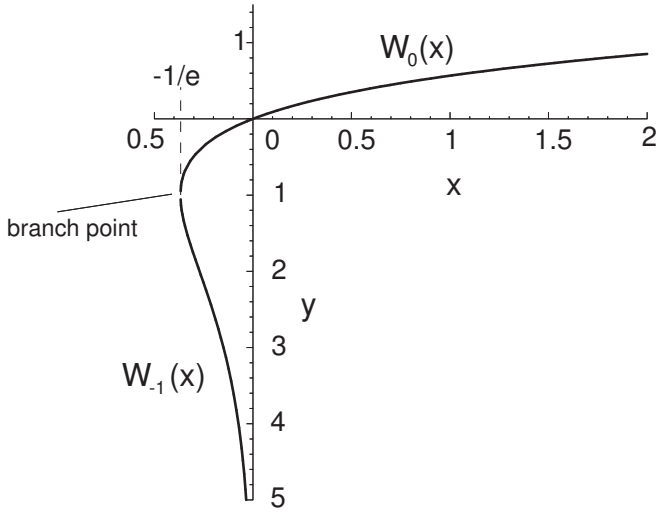}
\caption{The real branches $W_0(\xi)$ and $W_{-1}(\xi)$ of the 
  Lambert $W$-function.}
\label{f:lambert}
\end{figure}

Equations of the form of Eq.~(\ref{eq:lambrel}) appear in various
contexts such as enzymology, population growth processes and
hydrodynamics (see e.g. Ref.~\cite{corless_etal:96}). They are known
to have an explicit solution written in terms of a special function
called {\em $W$-function} \cite{corless_etal:96}:
\begin{eqnarray}
\label{eq:lambsol}
\sigma(x) & = & W \left(Y(x)\right) \, ,  
\qquad \sigma(x) > 0  \nonumber \\
\sigma(x) &= & W \left(-Y(x)\right) \,  ,  
\quad \;  \sigma(x) < 0 \, .
\end{eqnarray}
The Lambert $W$-function (see Fig.~\ref{f:lambert}) is a multi-valued
function with two real branches, which we refer to as $W_0(\xi)$ and
$W_{-1}(\xi)$. The branches merge at $\xi=-1/e$, where the Lambert
$W$-function takes the value $-1$.  The first branch, $W_0(\xi)$, is
defined for $\xi \ge -1/e $ ; it diverges at infinity
sub-logarithmically.  The second branch, $W_{-1}(\xi)$, is always
negative and defined in the domain $-1/e \le \xi \le 0$.  In the
vicinity of the point $\xi=-1/e$ the function $W(\xi)$ behaves like a
square root of $\xi$ since one gets $\partial_\xi W=W/[(1+W)\xi]$ by
the definition of the Lambert $W$-function, $W(\xi)\exp(W(\xi))=\xi$.

Using these properties of the Lambert $W$-function, the branch of $W$
is selected according to the value of the rescaled density $\sigma$.
For $\sigma \in [-2K/(K-1),-1]$ the relevant solution is $W_{-1}(-Y)$,
while for $\sigma \in [-1,0]$ one obtains $W_{0}(-Y)$. Finally, in the
interval $\sigma \in [0,2/(K-1)]$ one finds $W_{0}(Y)$.

The solutions are matched to the boundary conditions at the left and
right ends according to the entry or exit rates. The left and right
solutions, $\rho_\alpha(x)$ and $\rho_\beta(x)$, are then computed
from the expressions in Eqs.~(\ref{eq:sigma}) upon using the
coordinate transformation given by Eq.~(\ref{eq:changecoord}).
\begin{figure}[htbp!]
\includegraphics[width=\columnwidth]{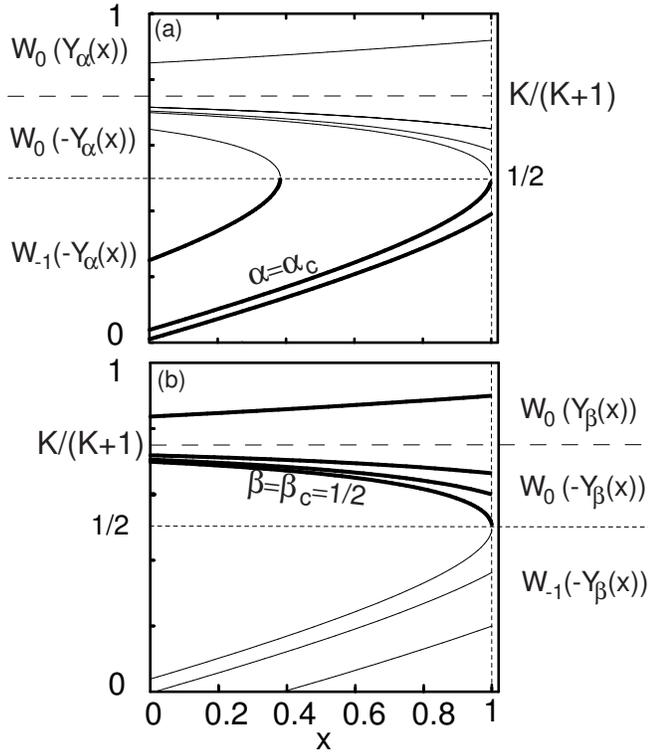}
\caption{Mathematical solutions for {\bf (a)} the left density
  $\rho_\alpha(x)$ and {\bf (b)} the right density $\rho_\beta(x)$ for
  $K=3$, $\Omega_D=0.1$ and different values of the entry and exit
  rate $\alpha$ and $\beta$.  The solutions which approach the
  Langmuir isotherm are those for $\alpha,\, \beta \le 1/2$ (thick
  lines). The solutions where the branching point coincides with the
  right boundary are indicated by $\alpha_c=0.038532...$ and
  $\beta_c=1/2$.}
\label{f:branches}
\end{figure}

Fig.~\ref{f:branches} provides a graphical representation of the
possible set of solutions of the first order differential equation,
Eq.~(\ref{eq:Bmod2}).  In order to decide which one of them are
actually physically realized, one needs to go back to the full
equation, either in its discrete form Eq.~(\ref{eq:Dnewmf}) or its
continuous version Eq.~(\ref{eq:mfa}). Analogous to the TASEP a
solution matching the density prescribed by the left boundary
condition is stable only if $\alpha < 1/2$~\footnote{For a fixed
  current $j$ the stationary density profile of the TASEP can be
  written as a non-linear map with $\rho_{i+1} = 1- j/\rho_i$.
  Consider a boundary value problem with a prescribed density $\rho(L)
  = 1 -\beta$ and $\beta \le 1/2$.  Then the attractive fixed point of
  the nonlinear map for forward iteration is $\rho=1-\beta \ge 1/2$.
  If now one starts with a density $\rho(0) = \alpha \ge \beta$ at the
  left end it will quickly converge towards the attractive fixed point
  of the non-linear map, $\rho = 1-\beta$; we then have a boundary
  layer at the left end.  The system is in the high density phase with
  bulk density and current determined by the right boundary.
  Otherwise, i.e. for $\alpha \le \beta$ and $\alpha \le 1/2$, a
  stable high density solution is not possible and both bulk current
  and density are determined by the left boundary.}. Such solutions
are shown as thick lines in Fig.~\ref{f:branches}(a). They are
monotonically increasing towards the Langmuir isotherm $\rho_l
=K/(K+1)> 1/2$.  This can be understood as a consequence of the
accumulation of particles from the bulk reservoir via the Langmuir
kinetics with increasing distance from the left boundary.  One might
now expect that the density will finally approach the Langmuir
isotherm. But, this is not the case. Instead, we find that the density
$\rho_\alpha(x)$ never increases beyond $1/2$, where the current
reaches its largest possible value $j_{\text{max}} = 1/4$.
Mathematically, this is a direct consequence of the analytic
properties of the Lambert $W$-function, which has a branching point at
a density $1/2$; see Fig.~\ref{f:branches}(a).  With decreasing
$\alpha$ the site where $\rho_\alpha (x)$ meets the density $1/2$
moves to the right. At a critical value of the entry rate, $\alpha_c
(\Omega_D, K)$, the branching point of the left solution $\rho_\alpha$
touches the right boundary.
 
Similarly to the discussion in the previous paragraph, solutions
matching the right boundary condition are stable only if $\beta \le
1/2$. The corresponding density profiles, shown as thick lines in
Fig.~\ref{f:branches}(b), are always in a high density regime, i.e.\/
$\rho_\beta(x) \ge 1/2$. If the density at the right boundary matches
the Langmuir isotherm, the right solution is flat $\rho_\beta(x)=
\rho_l$.  Otherwise, the source terms do not cancel, leading to a net
detachment/attachment flux such that the right density profiles decay
monotonically towards the Langmuir isotherm as one moves from the
right boundary to the bulk. As a consequence, the right density
$\rho_\beta(x)$ never crosses the Langmuir isotherm. The density
profile for $\beta=1/2$ is an {\em extremal solution} exhibiting the
lowest possible density ($\rho = 1/2$) and highest current ($j=1/4$)
at the right end, which then also coincides with the branching point
of the Lambert $W$-function.

In conclusion, for the left rescaled solution $\sigma_\alpha (x)$, an
entry rate $0 \le \alpha \le 1/2$ implies $-2K/(K-1) \le \sigma \le
-1$. Hence we have according to the previous analysis:
\begin{subequations}
\label{eq:sigma}
\begin{equation}
\label{eq:sigmaa}
\sigma_\alpha(x) = W_{-1}(-Y_\alpha (x)) < 0 \, .
\end{equation}
For the right rescaled solution $\sigma_\beta(x)$, one finds
correspondingly
\begin{equation}
\label{eq:sigmab}
\sigma_\beta(x) = \left\{
\begin{array}{lll}
W_{0}(Y_\beta(x)) > 0 & , \quad & 0 \le \beta < 1-\rho_l \\
& & \\
0 & , \quad  & \beta = 1-\rho_l \\
&  & \\
W_{0}(-Y_\beta(x)) < 0 &  , \quad & 1-\rho_l < \beta \le 1/2 \, ,
\end{array}
\right. 
\end{equation}
\end{subequations}
where $\rho_l=K/(K+1)$ is the constant density of the Langmuir
isotherm.  After the coordinate change (\ref{eq:changecoord}), the
general solution of the continuum mean-field equation at $\varepsilon
\rightarrow 0^+$, Eq.~(\ref{eq:mfa0}), is obtained by matching left
and right solutions $\rho_\alpha$ and $\rho_\beta$.  The remaining
task is now to identify the different scenarios where domain walls and
boundary layers appear.  Such analytic results are confirmed by the
numerical computation at finite $\varepsilon$.

\paragraph{Lower left quadrant: $\alpha, \beta \le 1/2$.}

This is the only case where there are solutions that approach the
Langmuir isotherm in the bulk and match both boundary conditions.  The
full density profile is obtained by finding the position $x_w$ where
the left and right currents coincide, i.e. $\rho_\alpha(x_w) =
1-\rho_\beta(x_w)$. One has to consider three cases: (i) $0 < x_w <
1$, (ii) $x_w < 0$, and (iii) $x_w > 1$.

\begin{figure}[htbp!]
\includegraphics[width=\columnwidth]{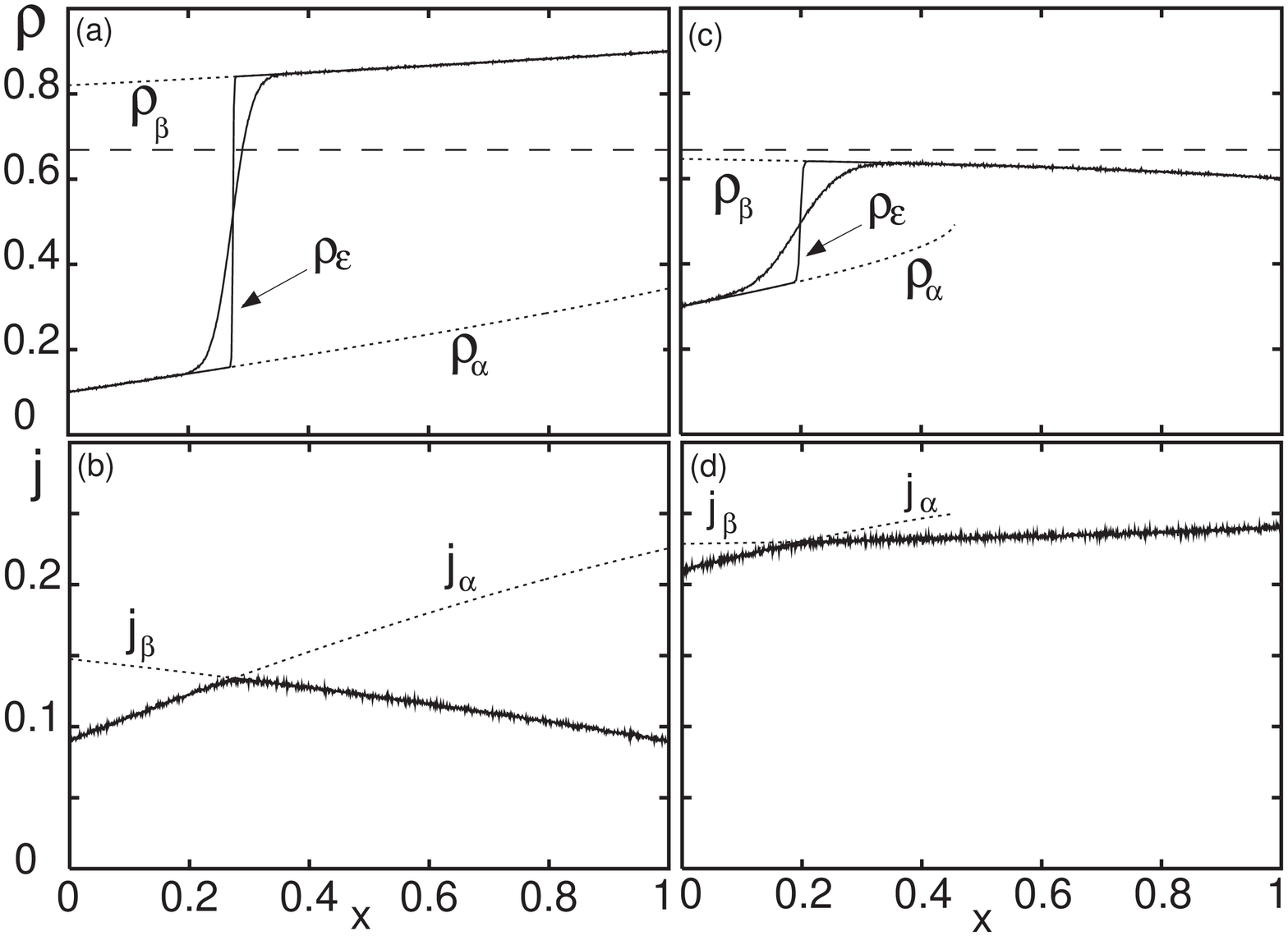}
\caption{Average density $\rho(x)$ {\bf (a)}-{\bf (c)} and corresponding 
  current $j(x)$ {\bf (b)}-{\bf (d)} for $\alpha,\, \beta \leq
  \frac12$ in a parameter regime showing phase separation. We have
  chosen $\Omega_D=0.1$, $K=2$ and (a)-(b) $\alpha=0.1,\,\beta=0.1$ or
  (c)-(d) $\alpha=0.3,\, \beta=0.4$. Solid lines correspond to the
  numerical solution of the mean-field theory with
  $\varepsilon=10^{-3}$.  Monte-Carlo simulations are shown as solid
  wiggly line. The flat dashed line represents the Langmuir isotherm,
  $\rho_l=K/(K+1)$.  The other dashed lines represent the analytic
  solutions given by the branches of the Lambert $W$-functions
  matching the boundary conditions on the right and left end,
  respectively.  For both cases (a) and (c), the solution matching the
  left boundary condition $\rho_\alpha$ is given by the branch of the
  Lambert $W$-function $W_{-1}(-Y_\alpha)$. For the solution matching
  the right boundary condition, $\rho_\beta$, one has to consider the
  branch $W_{0}$. For (a) the branch of $W$ has the argument
  $Y_\beta(x)$, while for (c) the argument is $-Y_\beta(x)$ (see as
  illustration also Fig.~\ref{f:branches}).}
\label{f:dr1}
\end{figure}

In case (i), a domain wall is formed separating a region of low
density on the left with a region of high density on the right.
Depending on whether $1-\beta$ is above or below $\rho_l$, different
profiles are observed, see Figs. (\ref{f:dr1})(a),(c). In the case
$\beta=1-\rho_l$, one obtains a flat profile of $\rho_\beta$ matching
the value of the Langmuir isotherm $\rho_l$. We note again that the
left and right solutions approach the Langmuir isotherm in the bulk.
In analogy with the case $K=1$, the domain wall is stabilized by the
current profiles controlled by the boundary conditions.

\begin{figure}[htbp!]
\includegraphics[width=0.7\columnwidth]{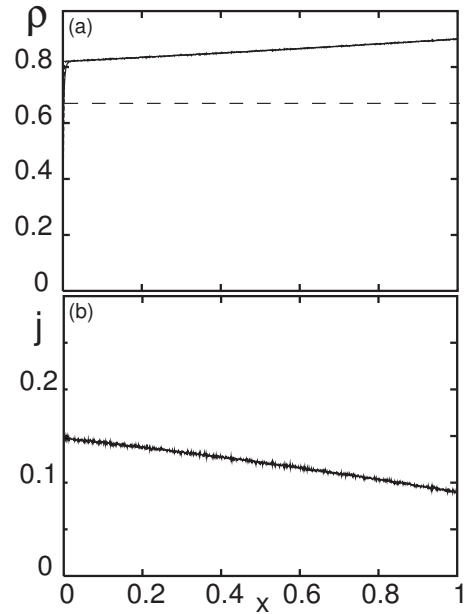}
\caption{Average density $\rho(x)$ {\bf (a)} and current $j(x)$ {\bf (b)}  for
  $\alpha=0.3$, $\beta=0.1$, $\Omega_D=0.1$ and $K=2$. We use the same
  legend as in Fig.~\ref{f:dr1}.  Except the left boundary layer, the
  bulk density profile is given by the Lambert $W$-function,
  $\rho_\beta=W_0(Y_\beta(x))$.}
\label{f:dr2}
\end{figure}

\begin{figure}[htbp!]
  \includegraphics[width=0.7\columnwidth]{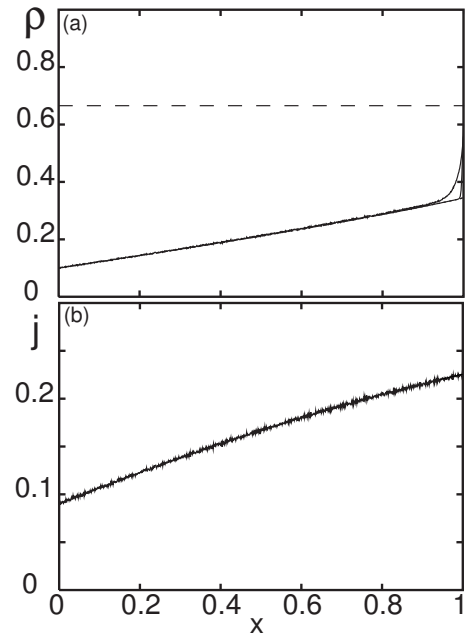}
\caption{Average density $\rho(x)$ {\bf (a)} and  current $j(x)$ {\bf (b)}  for
  $\alpha=0.1$, $\beta=0.4$, $\Omega_D=0.1$ and $K=2$. We use the same
  legend as in Fig.~\ref{f:dr1}.  Except the right boundary layer, the
  bulk density profile is given by the Lambert $W$-function,
  $\rho_\alpha=W_{-1}(Y_\alpha(x))$.}
\label{f:dr3}
\end{figure}

In cases (ii) and (iii), one of the two phases is
confined to the boundary.  Explicitly, for (ii) the bulk is
characterized by a HD with a boundary layer at the left end, see
Fig.~\ref{f:dr2}(a).
Correspondingly, for (iii) the solution exhibits a LD bulk phase
accompanied by a boundary layer on the right end side of the system,
see Fig.~\ref{f:dr3}(a).

\paragraph{The upper left quadrant, $\alpha < 1/2$, $\beta > 1/2$.}

As discussed above, for $\beta > 1/2$ the solutions of the first order
differential equation, Eq.~(\ref{eq:mfa0}), matching the right
boundary condition are physically unstable. Instead, the actual
density profile at the right boundary approaches the extremal solution
$W_0(-Y_{\beta=1/2})$ of the first order differential equation. The
density difference to the boundary value is bridged by a boundary
layer, which vanishes in the limit $\varepsilon \to 0^+$.

For the discussion of the density profiles in the upper left quadrant
we can simply parallel the arguments used for the lower left quadrant,
once the right solution has been substituted with the extremal one.
Depending on the matching of the current, one finds again three cases,
a LD phase, a 2-phase LD-HD coexistence and a HD phase.  We conclude
that the phases of the lower left quadrant extend to $\beta > 1/2$
with phase boundaries which are independent of the exit rate $\beta$,
i.e.  parallel to the $\beta$ axis. The HD phase for $\beta > 1/2$ has
some interesting features which are genuinely distinct from the HD
phase for $\beta < 1/2$. The density profile in the bulk is {\em
  independent} of the entrance and exit rates, $\alpha$ and $\beta$,
at the left and right boundaries; it is given by the extremal solution
$W_0 (- Y_{\beta=1/2} )$. The density approaches $\rho(L) = 1/2$ and
hence the current the maximal possible value $j_\text{max}=1/4$ at the
right boundary. These features are reminiscent of the maximal current
phase for the TASEP. The only difference seems to be that here current
and density are spatially varying along the system while they are
constant for the TASEP.  The essential characteristic in both cases is
that the behavior of the system is determined by the bulk and not the
boundaries. One is reminded of similar behavior of the Meissner phase
in superconducting materials. In the ensuing phase diagram we will
hence indicate this regime as the ``Meissner'' (M) phase to
distinguish it from the HD phase with boundary dominated density
profiles~\footnote{Though this behavior was mentioned explicitly in
  Ref.~\cite{parmeggiani_franosch_frey:03} we did not indicate it
  explicitly in the phase diagram.  Here we found this useful to
  emphasize the special nature of the high density phase above $\beta
  =1/2$. }.  Note also that the parameter range for the M phase is
broadened as compared to the maximal current phase of the TASEP.

\paragraph{The remaining quadrants, $\alpha > 1/2$.}

\begin{figure}[htbp!]
\includegraphics[width=0.7\columnwidth]{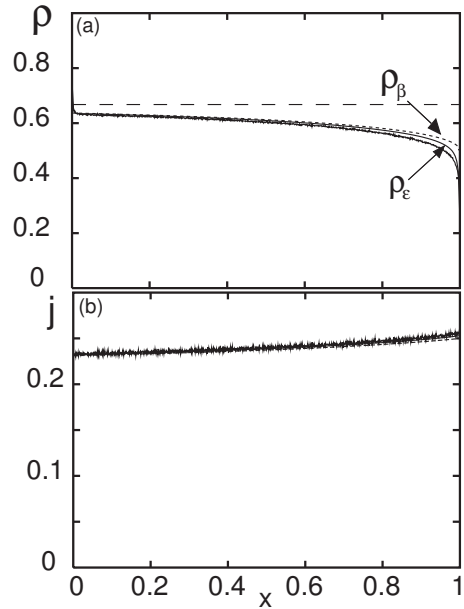}
\caption{Average density $\rho(x)$ {\bf (a)} and current $j(x)$ {\bf (b)}  for 
  $\Omega_D=0.1$, $K=2$, $\alpha=0.75$ and $\beta=0.75$. We use the
  same legend as in Fig.~\ref{f:dr1}.  Except for the left and right
  boundary layers, the bulk profile obtained from the analytic
  mean-field result is given by the branch
  $\rho_\beta(x)=W_0(-Y_\beta(x))$ of the Lambert $W$-function
  computed for $\beta=1/2$.} \label{f:dr5}
\end{figure}

At $\alpha=1/2$, the system is already in the high density phase where
the bulk profile does not match the entry rate.  Increasing $\alpha$
beyond the value $1/2$, therefore merely affects the boundary layer at
the left end.  The density profile is given by the right solution
$\rho_\beta$ for $\beta < 1/2$ or the extremal one for $\beta \ge 1/2$
as before; for an illustration compare Fig.~\ref{f:dr5}. For $\beta
\ge 1/2$ the same conclusion apply as in the preceding paragraph,
resulting in a ``Meissner'' phase for the upper right quadrant.

Let us conclude this subsection with some additional comments on
boundary layers.  Boundary layers arise from a mismatch between the
bulk profile and the boundary conditions. They can bend either upwards
or downwards depending on whether the left or right boundary rates are
above or below the values of the bulk solution at the ends.  For
example, in the right lower quadrant of the HD phase, a change from a
depletion to an accumulation layer at the left end of the system
occurs at $\alpha=\rho_\beta(0)$ for $\beta < 1/2$.
    
\subsubsection{The phase diagram}
\label{s:phasedr}

\begin{figure}[htbp!]
\includegraphics[width=\columnwidth]{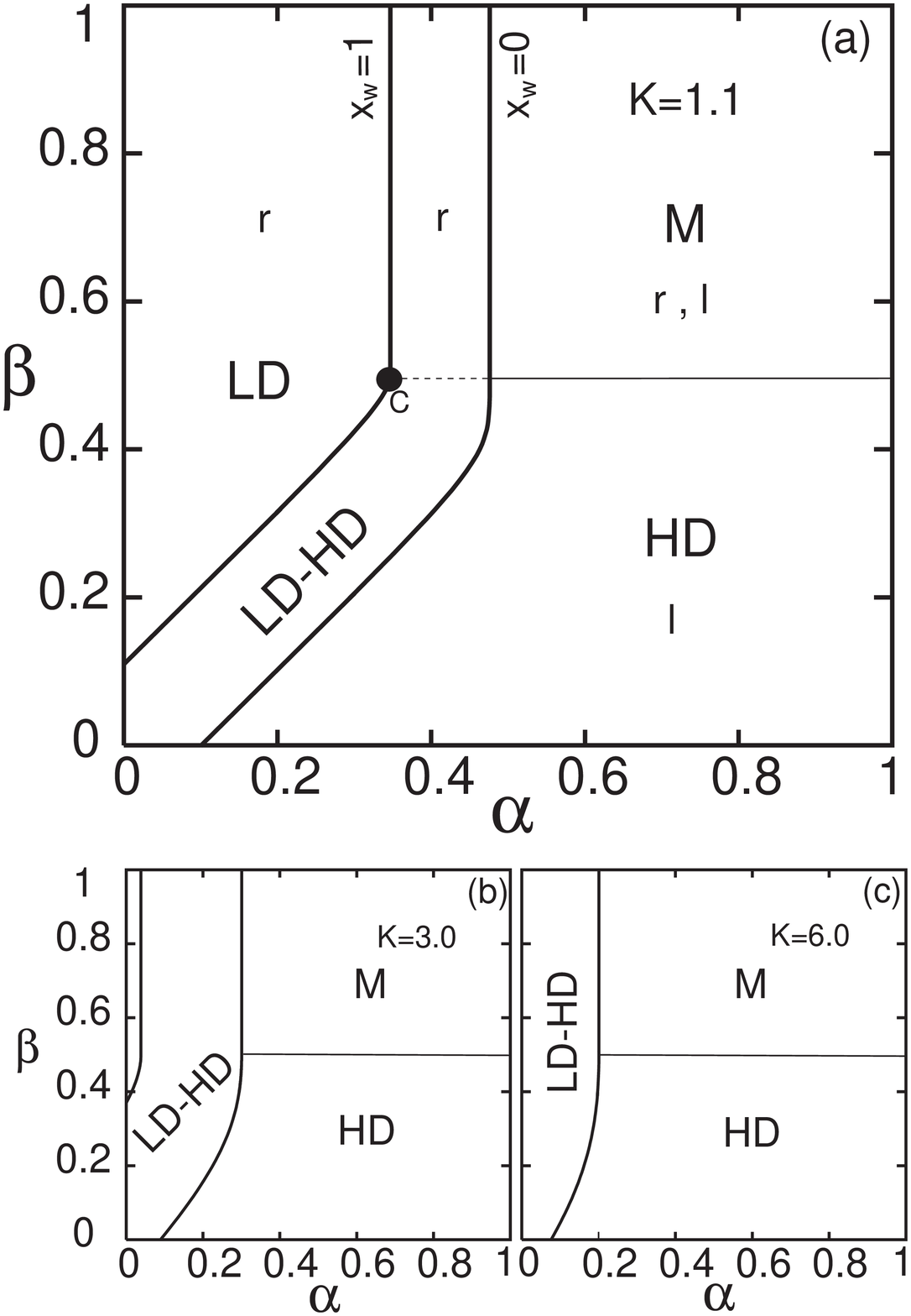}
\caption{Cuts of the phase diagrams on the $(\alpha,\, \beta)$--plane 
  obtained by the exact solution of the stationary mean-field equation
  (\ref{eq:mfa0}) in the inviscid limit $\varepsilon=0$ for
  $\Omega_D=0.1$ and {\bf (a)} $K=1.1$, {\bf (b)} $K=3.0$, {\bf (c)}
  $K=6.0$. The two lines, corresponding to regions in which the domain
  wall is located at $x_w=0$ and $x_w=1$, are obtained by using the
  matching conditions for the currents: $j_\alpha(1)=\beta(1-\beta)$
  and $j_\beta(0)=\alpha(1-\alpha)$. In figure (a), we emphasize
  several features.  With the letters "l" and "r" we indicate the
  presence of boundary layers in the average density profile, forming
  at the left or the right end of the system, respectively. In the
  lower left quadrant, the left and right boundary layers form
  whenever the domain wall exits the system on the left and right end
  side.  At the phase boundary between the HD and M phases, for
  $\beta=1/2$, a boundary layer forms at the right end. Note that also
  in the M phase $\rho_{bulk} > 1/2$. The presence of boundary layers
  in the different phases of the $(\alpha,\, \beta)$--plane is
  conserved upon variation of the binding constant $K$.  The filled
  black circle represents the critical point $C$ where the domain wall
  exhibits critical behavior; see Sect.~\ref{s:dwp}. This critical
  point exits the plane for large values of $K$, accompanied with a
  topological change of the phase diagram.}
\label{f:phaseKdr}
\end{figure}

We discuss the topology of the phase diagram with respect to cuts in the
$(\alpha,\beta)$--plane for different values of $K$ and $\Omega_D$.
We first consider the situation in which $\Omega_D$ is fixed and $K$
increases, starting from values slightly larger than unity. Figure
\ref{f:phaseKdr}(a) shows the phase diagram for $K=1.1$.  A low
density (LD) phase occupies the upper left of the plane, while a high
density (HD) and a Meissner (M) phase are located on the right. In
between there is a 2-phase coexistence region (LD-HD).  In the
coexistence phase a domain wall is localized at the point $x_w$ in the
bulk.  The boundaries of the coexistence region in the phase diagram
are determined by those parameters where the domain wall hits either
the entrance, i.e. $x_w=0$, or the exit of the system, $x_w=1$.  For
$\beta > 1/2$, the density profile only develops a boundary layer at
the right end, but remains unchanged in the bulk.  Since the domain
wall position becomes independent of $\beta$, the boundaries of the
2-phase coexistence region become parallel to the axis $\alpha=0$.  It
is important to remark that from the analytic results the left
solution $\rho_\alpha$ is strictly smaller than $1/2$, except for the
special point $C$ in the phase diagram where $\rho_\alpha(1) = \beta =
1/2$.  We shall see in Section \ref{s:dwp} that in the vicinity of
this point the domain wall exhibits critical properties.

Upon increasing $K$, the LD phase progressively shrinks to a region
close to the $\beta$-axis, while the size of the two other phases
increases; see Figs. \ref{f:phaseKdr}(b) and \ref{f:phaseKdr}(c). A change of topology occurs when the LD phase collapses on
this axis which happens upon passing a critical value of $K$. This
critical value depends on $\Omega_D$ and can be computed using the
expressions in Eqs.  (\ref{eq:dwp1}) and (\ref{eq:sigma}).  A further
increase of $K$ results in a decrease of the extension of the LD-HD
region in the phase diagram, see Figs.  \ref{f:phaseKdr}.  Eventually,
for very large $K$ the average bulk density in the HD and M regions
approaches saturation $\rho_{bulk}=1$.

\begin{figure}[htbp!]
\includegraphics[width=\columnwidth]{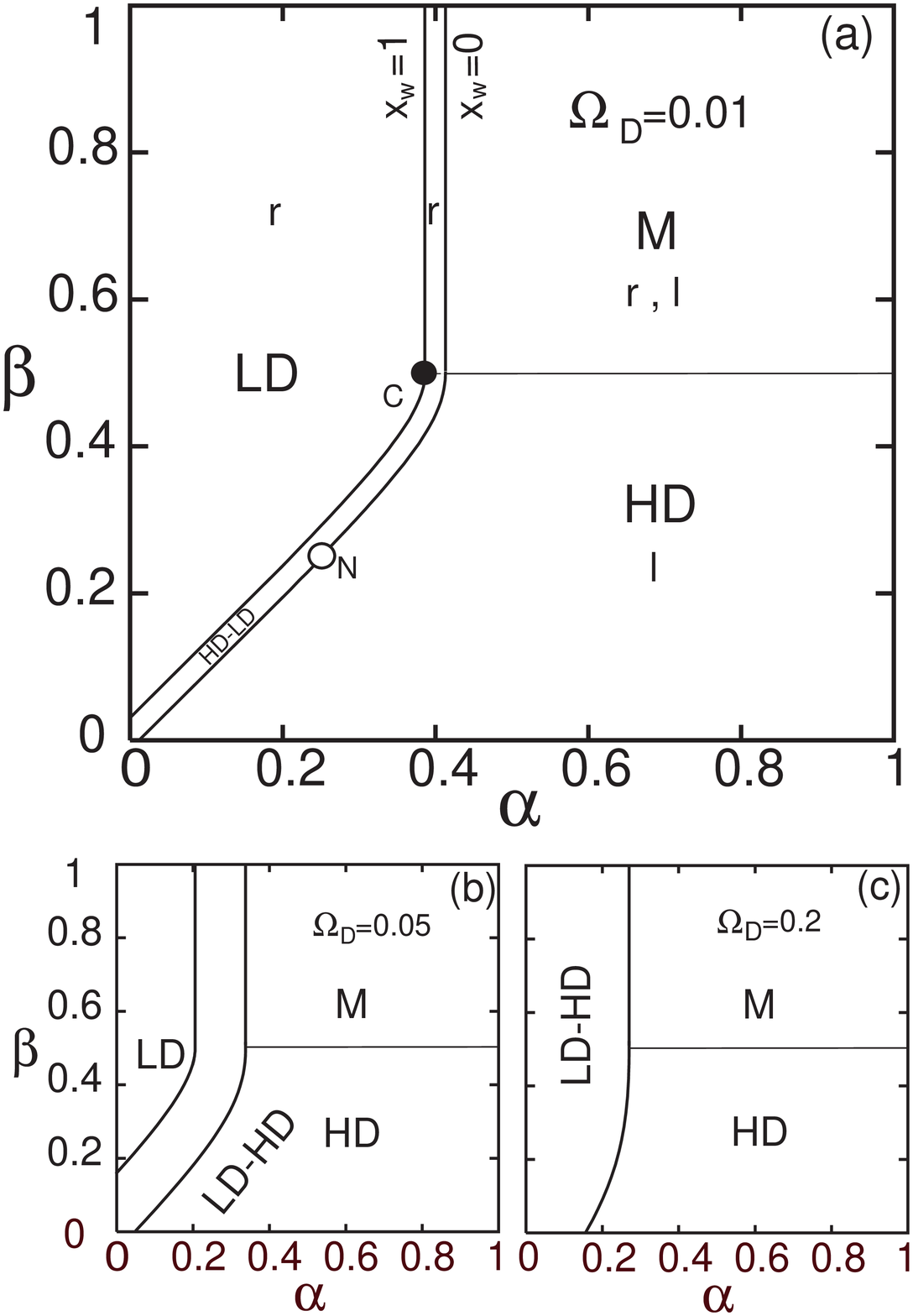}
\caption{Cuts of the phase diagrams as in Fig.~\ref{f:phaseKdr} for $K=3$ and
  {\bf (a)} $\Omega_D=0.01$, {\bf (b)} $\Omega_D=0.05$, {\bf (c)}
  $\Omega_D=0.2$.  The white circle corresponds to a {\em nodal} point
  of the system ${\cal N}$ defined by the condition
  $\alpha=\beta=1-\rho_l=1/(K+1)$.  Every line $x_w=0$ crosses this
  point for an increasing $\Omega_D$.}
\label{f:phaseOdr}
\end{figure}

Similarly, increasing $\Omega_D$ at fixed $K$, the same topology
change occurs, as described above; see Fig. \ref{f:phaseOdr}.  However, we note the different
limiting behaviors for $\Omega_D \to 0^+$ and $\Omega_D \to \infty$.
In the first case, we are considering the limit of the model to the
TASEP for a given binding constant $K$ (although $K\ne 1$). The
2-phase coexistence region LD-HD shrinks continuously to the line
$\alpha=\beta$.  In the same limit, in the upper right quadrant,
$\alpha,\beta > 1/2$, the M phase approaches continuously the MC phase
of the TASEP. For a very large detachment rate $\Omega_D$, the right
boundary of the LD-HD coexistence phase approaches a straight line at
finite entry rate $\alpha$ that can be computed from the analytic
solution as equal to $1 - \rho_l$. In the same limit, the average
density in the bulk reaches asymptotically the value $\rho_{bulk} =
\rho_l$ of the Langmuir isotherm.

Eventually, one observes that all phase boundaries between the LD-HD
coexistence and the HD phase, i.e. where the domain is pinned at $x_w
= 0$, intersect at the same point ${\cal N}$ for any value of the
detachment rate $\Omega_D$.  This {\em nodal} point ${\cal N}$ can be
evaluated as $\alpha = \beta = 1-\rho_l = 1/(K+1)$. At this point,
indeed, the average density $\rho$ is given by the flat profile of the
Langmuir isotherm $\rho_l = K/(K+1)$ which is obviously independent of
$\Omega_D$. As a result, the domain wall does not move from $x_w = 0$
for any value of the detachment rate $\Omega_D$. Interestingly, one
remarks that both points $C$ and ${\cal N}$ approach continuously the
{\em triple} point of the TASEP $\alpha = \beta = 1/2$ in the
simultaneous limit $\Omega_D \to 0^+$ and $K \to 1^+$.

\section{Domain wall properties}
\label{s:dw}

The knowledge of the analytic solution in the mean-field approximation
allows for a detailed study of the behavior of the domain wall height
and position upon a change of the system parameters. While the results
for the symmetric case $K=1$ are more or less trivial, novel
properties emerge for $K>1$.  In this Section, we shall start from the
description of the domain wall behavior on the $(\alpha,\beta)$--plane
of the phase diagram along trajectories of constant entry or exit
rates, respectively.

\subsection{Position and amplitude of the domain wall on the 
  $(\alpha,\beta)$--plane}

\begin{figure}[htbp!]
\includegraphics[width=\columnwidth]{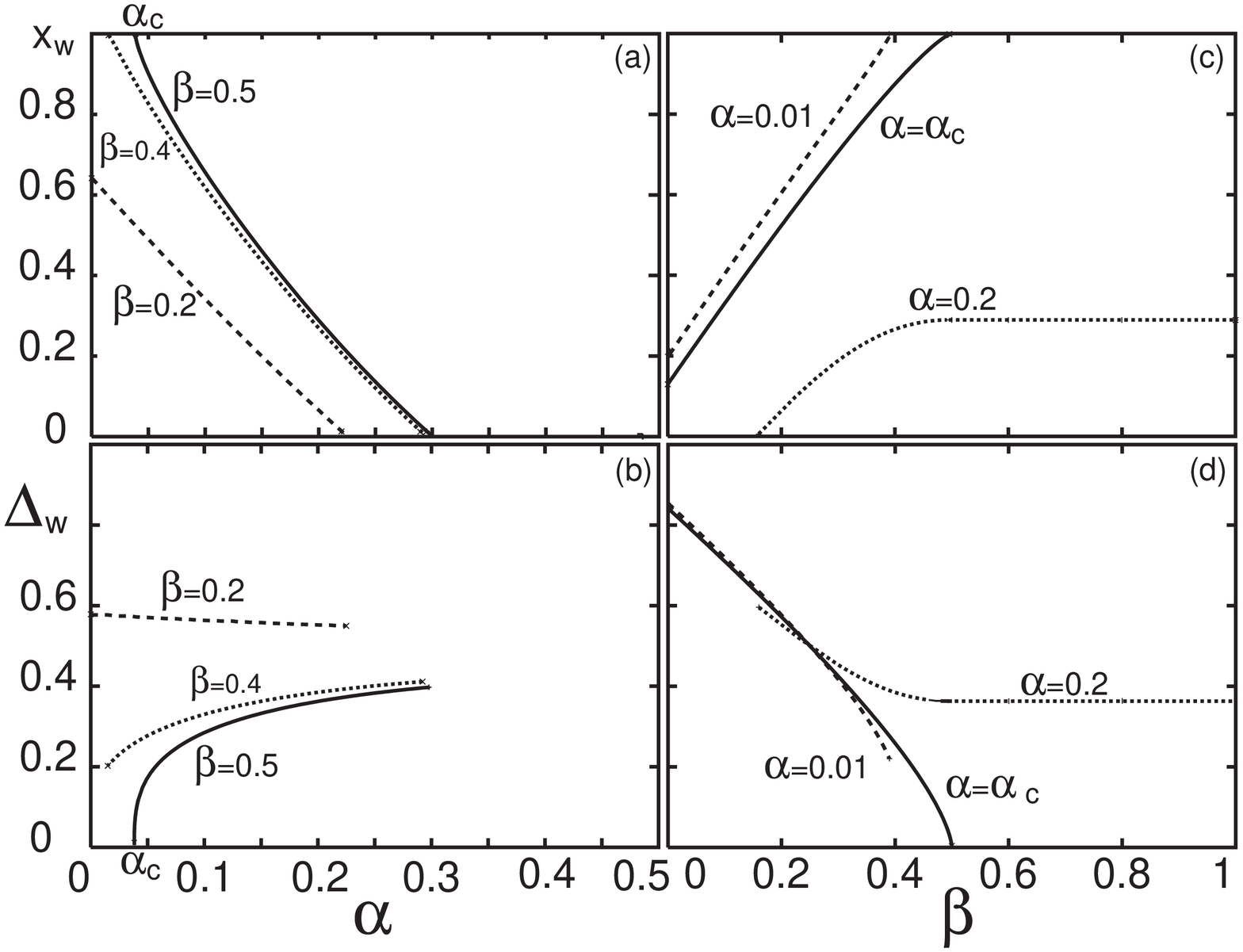}
\caption{{\bf (a)}-{\bf (b)} Domain wall position $x_w$ and height 
  $\Delta_w$ as a function of the entrance $\alpha$ for
  different values of the exit rate $\beta$ at $\Omega_D=0.1$ and
  $K=3$. At the critical point $\alpha=\alpha_c$ and $\beta=1/2$ a
  domain wall forms at the right end of the system with an
  infinitesimal height $\Delta_w$. The value of the "critical" entry
  rate is $\alpha_c=0.038532...$ and can be written explicitly by
  using the analytic solution in the mean-field approximation, see
  Eq.~(\ref{eq:alphacritical}).
  {\bf (c)}-{\bf (d)} Domain wall position $x_w$ and height 
  $\Delta_w$  as a function of the exit rate $\beta$ for
  different values of the entrance rate $\alpha$ at $\Omega_D=0.1$ and
  $K=3$.  For $\alpha=\alpha_c$ and $\beta=1/2$ a domain wall forms at
  the right end of the system with an infinitesimal height $\Delta_w$.
  For exit rates $\beta > 1/2$, both domain wall position $x_w$ and
  height $\Delta_w$ become independent of $\beta$. Changes in the exit
  rate only affect the size and shape of the boundary layer on the
  right end, but not the bulk density profile.}
\label{f:DWalpha-DWbeta}
\end{figure}

Figures \ref{f:DWalpha-DWbeta}(a,b) show the dependence of the domain wall
position, $x_w$, and height, $\Delta_w$, on the entry rate $\alpha$
along lines of constant exit rate $\beta$. As can be inferred from the
structure of the phase diagram presented in the preceding Section, for
a small enough exit rate $\beta$, a domain wall can form in the bulk
with a finite amplitude even for a vanishing entry rate, $\alpha=0$.
For larger $\beta$, one observes that the domain wall builds up with a
finite height on the right boundary only above some specific value of
$\alpha$. If one regards the domain wall height as a kind of order
parameter for the coexistence phase such a behavior can be termed a
first order transition. This has to be contrasted with the case
$\beta=0.5$, where the domain wall enters the system at $x_w=1$ with
infinitesimal height at a critical entry rate $\alpha=\alpha_c$. In
the same terminology this would then be a second order transition.
Indeed, as we are going to discuss in the next subsection, the domain
wall exhibits critical properties at this point.  In the phase diagram
(Fig.~\ref{f:phaseKdr}(a)) the corresponding critical point is
indicated as $C$.

In all cases, upon increasing $\alpha$ and hence the influx of
particles, the domain wall changes its position continuously from the
right to the left end of the system. Then, at some value $\alpha$
which depends on $\beta$, the domain wall leaves the system with a
finite amplitude $\Delta_w$.

Similar behaviors of the position and height of the domain wall is
found as a function of $\beta$ for fixed values of $\alpha$; see
Fig.~\ref{f:DWalpha-DWbeta}(c,d).  Here one finds that, upon increasing $\beta$ and
hence reducing the out-flux of particles the domain wall position
$x_w$ moves continuously from the left to the right boundary .  For
small $\alpha$, a domain wall is formed at a finite position $x_w$ and
$\beta = 0$. For larger entry rates, the domain wall forms at $x_w =
0$ with a finite amplitude only for finite values of the exit rate
$\beta$. As before, the amplitude of the domain wall $\Delta_w$
vanishes only for the critical value $\alpha = \alpha_c$ at $\beta =
1/2$. Indeed, when $\alpha > \alpha_c$ and $\beta > 1/2$, one notes
that the domain wall position $x_w$ remains constant upon changing
$\beta$.  As we have explained above, this corresponds to the
situation where the bulk profile is unaffected by a change in the exit
rate (M phase). Only the magnitude of the boundary layer changes with
increasing $\beta$.

\subsection{Critical properties of the domain wall}
\label{s:dwp}

In this Section, we discuss the domain wall properties close to the
special point $C$ where the domain wall forms with infinitesimal
height. The analysis will make use of the analytic solution in the
mean-field approximation. We show that the domain wall emerges as a
consequence of a bifurcation phenomenon, and calculate the resulting
non-analytic behavior of its height and position.

\begin{figure}[htbp!]
\includegraphics[width=\columnwidth]{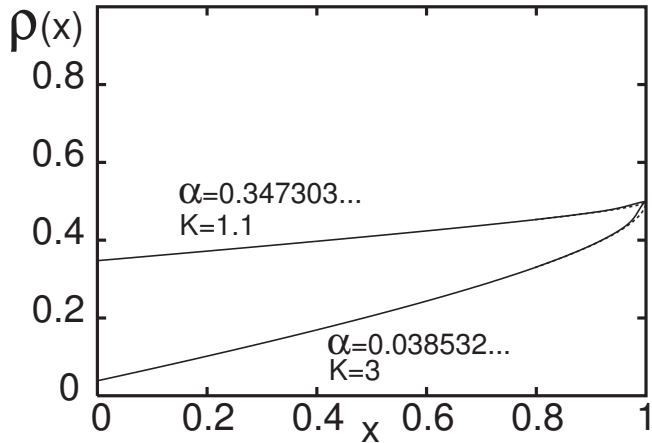}
\caption{Average density profiles computed analytically in the
  inviscid limit expressed in terms of Lambert $W$-function (dashed
  lines) and numerically for $\varepsilon=10^{-3}$ within a mean-field
  approximation (solid smooth line).  Parameters are: $\alpha =
  \alpha_c$ (see Eq.~(\ref{eq:alphacritical}) for the analytic
  expression), $\beta = 0.5$, $\Omega_D = 0.1$ and different values of
  $K$. The profile is entirely given by the left solution
  $\rho_\alpha$ for the value of the entry rate $\alpha_c$, defined by
  the condition $\rho_\alpha(x=1) = 1/2$, and $\beta = 1/2$. Note that
  in this case, $\rho_\alpha$ matches simultaneously the left and the
  right boundary conditions.}
\label{f:dr6}
\end{figure}

At the point $C$, the analytic solution of the mean-field equations is
described by a low density profile $\rho(x) = \rho_\alpha(x)$ that not
only matches the boundary conditions at the left but also the one at
the right end; see Fig.~\ref{f:dr6}.
This implies that the left and right currents also match at the right
end of the system. Interestingly, at this position the current $j$ is
maximal~\footnote{Note that in a pure TASEP the point of the phase
  diagram that satisfies such properties is $\alpha = \beta = 1/2$.}.
By a small change of the system parameters in the 2-phase LD-HD
coexistence region the domain wall forms at the right end
characterized by a small height, provided that $\beta \le 1/2$.

Using the analytic solution (\ref{eq:sigma}) one can give an explicit
expression of the critical point $C$ as a function of $\Omega_D$ and
$K$. The condition that the left boundary matches the value $1/2$ at
$x = 1$ translates to $\sigma_\alpha(x=1) = -1$ or, using the
properties of the Lambert $W$-function, as:
\begin{equation}
\label{eq:criticalDW} Y_\alpha(x=1) = 1/e \, .
\end{equation} 
{F}rom the expression of the function $Y_\alpha$,
Eq.~(\ref{eq:gencoord}), and the initial condition $\sigma(0) =
(2\alpha-1)(K+1)/(K-1)-1$, one computes the value of the {\em
  critical} entry rate
\begin{equation}
\alpha_c \! =
\!\! \frac{K}{(K+1)} + 
\frac{K-1}{2(K+1)}W_{-1}\!\!\left(\!\!-\exp\!\left\{\!\!-\Omega_D
\frac{(K+1)^2}{(K-1)}\!-\!1\!\right\}\!\!\right)\!\! \, .   
\label{eq:alphacritical}
\end{equation}
{F}rom the discussion of the phase diagram and the general properties
of the domain wall, one already infers that $0 < \alpha < 1/2$ for not
too large values of $\Omega_D$ and $K$.\\
The set
\begin{equation}
\label{eq:criticalmanifold}
(\alpha=\alpha_c(K,\Omega_D),\, \beta=1/2,\, K,\, \Omega_D)
\end{equation}
defines a two dimensional smooth manifold in parameter space ({\em
  critical manifold}).
 
In order to study the critical properties close to this manifold we
apply standard methods of bifurcation theory~\cite{arnold:book,
  wiggins:book, guckenheimer_holmes:book}. We consider a smooth path
in the region of parameter space close to the critical manifold
defined above.  At some point $C$ this path will cross the critical
manifold.  The small quantities that describe the behavior of the
domain wall close to the critical manifold are the distance from the
right end side, $\delta x_c \equiv 1-x_w$, and the domain wall height,
$\Delta_w = \rho_\beta(x_w)-\rho_\alpha(x_w)$.  These quantities will
be expressed to leading order in terms of the small deviations from
the critical point $\delta\alpha=\alpha-\alpha_c$,
$\delta\beta=\beta-1/2$, and similarly for $\delta \Omega_D$ and
$\delta K$.

The matching condition of the left and right currents,
$\rho_\alpha(x_w) + \rho_\beta(x_w) = 1$, can be rewritten in terms of
reduced densities $\sigma$ as $\sigma_\alpha(x_w) + \sigma_\beta(x_w)
= -2$. As a consequence, the solution close to the critical point
writes as $\sigma_{\alpha,\beta}(x_w) = -1\mp \Delta\sigma$, where we
have introduced the reduced domain wall height $\Delta\sigma$ as
another small quantity.  The relation between $\delta x$ and
$\Delta\sigma$ can be obtained by expanding the equality
\begin{equation}
\label{eq:relation1}
\sigma_\beta \exp{(\sigma_\beta)}= -Y_\beta(x_w)  \, ,
\end{equation}
leading to
\begin{equation}
\label{eq:expansion1}
\delta x \sim (\Delta\sigma)^2 \, ,
\end{equation}
where the prefactor can be explicitly computed and depends only on the
value of the system parameters at the critical point $C$. A second
relation connecting $\Delta\sigma$ to the small distances
$\delta\alpha,\, \delta\beta,\, \delta K$ and $\delta\Omega_D$ arises
from the definition of the Lambert $W$-function
$|\sigma|\exp(\sigma)=Y$ by taking the ratio
\begin{equation}
\label{eq:relation2}
\frac{\sigma_\beta}{\sigma_\alpha} \exp{(\sigma_\beta-\sigma_\alpha)} = 
\frac{Y_\beta(x_w)}{Y_\alpha(x_w)}  \, .
\end{equation}  
The important observation is that the right-hand side is independent
of the domain wall position $x_w$; see Eq.~(\ref{eq:gencoord}).
Expanding Eq.~(\ref{eq:relation2}) to leading order, one obtains:
\begin{equation}
\label{eq:expansion2}
(\Delta\sigma)^3 \sim \delta O 
= A_\alpha \delta \alpha + A_\beta (\delta\beta)^2 + 
  A_K \delta K + A_\Omega \delta \Omega_D \, ,
\end{equation}
where $\delta O$ is a distance along a generic path that ends on the
critical manifold. We do not consider the non-generic case where the
the critical manifold is approached tangentially. Then one finds power
laws different from those presented below for the generic case.

\begin{figure}[htbp!]
\includegraphics[width=\columnwidth]{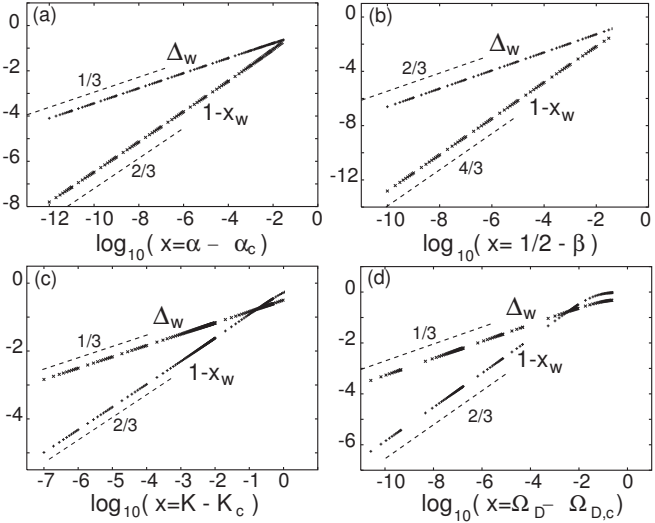}
\caption{Double decimal logarithmic plots of the critical behavior of
  the domain wall height, $\Delta_w$, and position from the right end
  side, $1-x_w$. We obtained the plot numerically with the program
  Maple, release 7, using the analytic mean-field solution in the
  vicinity of the critical point $C$ and applying the matching
  condition over the left and right currents, $j_\alpha(x_w) =
  j_\beta(x_w)$.  {\bf (a)} As a function of $\alpha$ starting from
  the point $C$ on the critical manifold with coordinates $\alpha_{c}
  = 0.038532...$, $\beta = 1/2$, $\Omega_D = 0.1$ and $K = 3$. {\bf
    (b)} As a function of $\beta$ starting from the point $C$ on the
  critical manifold with coordinates $\alpha = 0.038532...$, $\beta_c
  = 1/2$, $\Omega_D=0.1$ and $K=3$. {\bf (c)} As a function of $K$
  from the point $C$ on the critical manifold with coordinates
  $\alpha=0.2$, $\beta = 1/2$, $\Omega_D = 0.051443...$ and $K_c = 3$.
  {\bf (d)} As a function of $\Omega_D$ from the point $C$ on the
  critical manifold with coordinates $\alpha = 0.2$, $\beta = 1/2$,
  $\Omega_{D,c} = 0.051443...$ and $K=3$. The value of $\Omega_{D,c}$
  can be easily obtained from Eq.~(\ref{eq:gencoord}) and the initial
  condition $\sigma(0)$. Note the different scaling regime for the
  exit rate $\beta$.}
\label{f:powerlaw}
\end{figure}

As before, the coefficients $A$ can be computed explicitly and shown
to depend only on the rates at the critical point $C$.  Interestingly,
the distance $\delta O$ does not exhibit a linear term in $\delta
\beta$.  This is due to the singular behavior of the density profile
$\rho_\alpha(x)$ close to the right boundary at the critical point
$C$; see Fig.~\ref{f:dr6}.  Combining the two expansion, one finds the
following power laws:
\begin{eqnarray}
\label{eq:powerlaws}
\delta x \sim \delta O^{2/3} & \qquad & \Delta_w \sim \delta O^{1/3} \, .
\end{eqnarray}
The validity of these exponents is confirmed numerically in Figs.
\ref{f:powerlaw}.
We also checked that the amplitudes in the expansions
(\ref{eq:powerlaws}) coincide with those obtained by the numerical
data.

\subsection{Further properties of the domain wall position}

\begin{figure}[htbp!]
\includegraphics[width=\columnwidth]{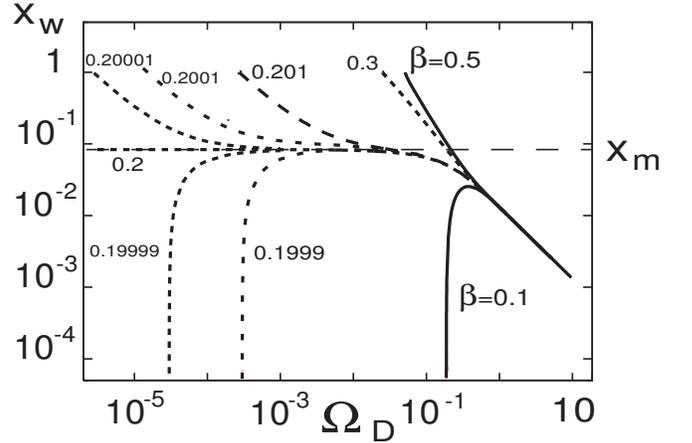}
\caption{Domain wall position $x_w$ in logarithmic scale as a function 
  of $\Omega_D$ at $\alpha=0.2$, $\Omega_D=.051443...$ and $K=3$ and
  different values of $\beta$. If $\beta > \alpha$ the domain wall
  builds up from the right boundary, while for $\beta<\alpha$ from the
  left boundary. For $\alpha = \beta$ the domain wall approaches the
  position $x_m$ which is independent of the decreasing detachment
  rate $\Omega_D$. At large $\Omega_D$ the domain wall position $x_w$
  always moves to the left boundary as $1/\Omega_D$.}
\label{f:DWposition}
\end{figure}

In this Section we discuss how the position of the domain wall $x_w$
moves upon changing $\Omega_D$ for fixed $\alpha$ and $K>1$ and a set
of different values for $\beta$.  In the first quadrant, $\alpha,\beta
< 1/2$, and for very small values of $\Omega_D$ the coexistence phase
is confined to a narrow strip parallel to the diagonal $\alpha =
\beta$; see Fig.~\ref{f:phaseOdr}(a). It extends to the quadrant
$\alpha < 1/2, \beta > 1/2$, where boundary layers form.  On the other
hand, for very large $\Omega_D$ the coexistence phase corresponds to
the region $\alpha < 1-\rho_l$, see Sect.  \ref{s:phasedr}. The
interesting features therefore arise in the region of $\alpha <
1-\rho_l$ and $\beta < 1/2$.  We consider a path in the phase diagram
with fixed $\alpha, \beta$ and $K$ and follow how it intersects the
phase boundaries of the 2-phase coexistence region LD-HD as $\Omega_D$
is increased.  From Fig.~\ref{f:DWposition} one can distinguish three
cases.

For $\alpha < \beta$ the system is in a LD phase for very small
$\Omega_D$. Then, at a critical value of $\Omega_D$ it enters the
LD-HD region where a domain wall forms at the right boundary. A
further increase of $\Omega_D$ results in a change of the domain wall
position to the left, asymptotically reaching the left boundary for
very large values of the detachment rate $\Omega_D$.

For $\beta < \alpha $, the system is in the HD region for small
$\Omega_D$. By increasing the detachment rate, it enters the LD-HD
region. Differently from the previous case, the domain wall now forms
at the left boundary, it moves to the right up to a maximal position
$x_m$ for intermediate values of $\Omega_D$, and finally for large
$\Omega_D$ it moves back to the left boundary with the same asymptotic
behavior as the previous case.

For $\beta = \alpha$, the system remains in the 2-phase coexistence
region LD-HD for all values of the detachment rate $\Omega_D$. One can
prove, using the analytic solution (\ref{eq:sigma}), that the domain
wall position stays finite even in the limit $\Omega_D \to 0^+$ and is
given by
\begin{equation}
\label{eq:DWposition}
x_w = \frac{\sigma(1)(1+\sigma(0))}{2(\sigma(0)\sigma(1)-1)} \, ,
\end{equation}    
where $\sigma(0)$ and $\sigma(1)$ are the usual boundary conditions
written in terms the model parameters; see Eq.~\ref{eq:changecoord}.
Interestingly, the domain wall position $x_w$ obtained for $\alpha =
\beta$ and vanishing $\Omega_D$ does not reduce to the value given by
the mean-field continuum approximation in a pure TASEP, i.e.
$x_w=1/2$.  In order to regain the TASEP position, the binding
constant $K$ has to approach the unity. Moreover, from
Eq.~(\ref{eq:DWposition}) one find that in the limit $\alpha = \beta
\to 1/2^-$, the position $x_m$ is a singular function of the binding
constant close to $K=1^+$.

The previous discussion corroborates the fact that the Langmuir
Kinetics constitutes a singular perturbation of the TASEP even in the
limit of small rates, yielding new features that are generated by the
competition between the two dynamics.

\section{Conclusions}
\label{s:conclusions}

We have presented a detailed study of a model for driven
one-dimensional transport introduced recently in
Ref.~\cite{parmeggiani_franosch_frey:03}, where the dynamics of the
totally asymmetric simple exclusion process has been supplemented by
Langmuir kinetics. This non-conservative process introduces
competition between boundary and bulk dynamics suggesting a new class
of models for non-equilibrium transport.  The model is inspired by
essential properties of intracellular transport on cytoskeletal
filaments driven by processive motor
proteins~\cite{alberts,howard:book}.  These molecular engines move
unidirectionally along cytoskeletal filaments, and simultaneously are
subject to a binding/unbinding kinetics between the filament and the
cytoplasm. The processivity of the motors implies low rates of
detachment. Attachment rates can be easily tuned by changing the
concentration of motors in the cytoplasm.  In particular one may
obtain very low attachment rates using a low volume concentration of
motors.

The non-conservative dynamics proposed introduces a non-trivial
stationary state, with features qualitatively different from both the
totally asymmetric simple exclusion process and Langmuir kinetics.
The competing dynamics leads to an unexpected spatial modulation of
the average density profile in the bulk. For extended regions in
parameter space, we find that the density profile exhibits
discontinuities on the scale of the system size which is
characteristic for phase separation. Furthermore, the coexisting
phases manifest themselves by a domain wall that, contrary to the
TASEP, is localized in the bulk. In contrast to previous models
\cite{kolomeisky:98, mirin_kolomeisky:03}, the localization is not
induced by local defects, but arises via a collective phenomenon based
on a microscopically homogeneous bulk dynamics. The resulting phase
diagram is topologically distinct from the totally asymmetric
exclusion process and exhibits new phases.

An analytic solution for the density profile has been derived in the
context of a mean-field approximation in the continuum limit. The
properties of the average density for different kinetic rates are
encoded in the peculiar features of the Lambert $W$-function. In
particular, the discovery of a branching point is a prerequisite to
rationalizing the behavior of the solution. The analytic solution has
allowed to trace and study in detail the properties of the phase
diagram. We found that the cases of equal and different binding rates
give rise to rather distinct topologies in the phase diagram. The
limiting cases for small or large kinetic rates have been computed
analytically. We have identified special points of the phase diagram
which are the analogue of the "triple point" (viz. where all three
phase boundaries meet) of the totally asymmetric simple exclusion
process. There, a domain wall builds up with infinitesimal height at
the boundary, exhibiting critical features characterized by unusual
mean-field exponents. Finally, we have discussed some limiting cases
in which the properties of the totally asymmetric simple exclusion
process in the mean-field approximation are recovered.
 
\begin{figure}[htb]
\includegraphics[width=\columnwidth]{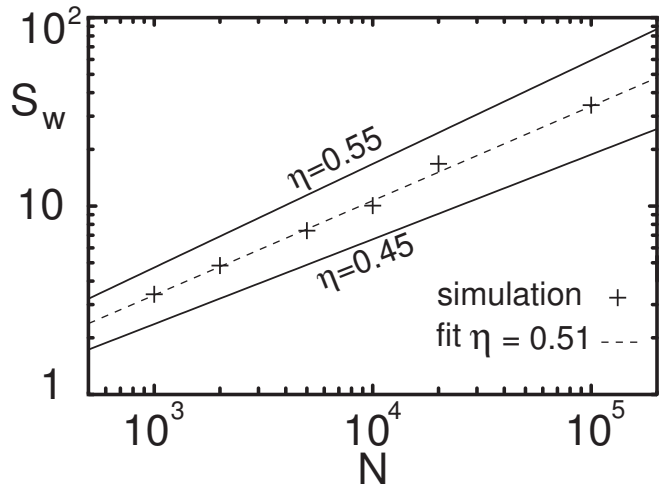}
\caption{Domain wall slope $S_w$ estimated from Monte 
  Carlo simulations as a function of the system size $N$.  Simulations
  where performed for $\alpha=0.2$, $\beta=0.6$, $K=3$
  and $\Omega_D=0.1$.}
\label{f:slope}
\end{figure}

Let us give some more intuitive arguments on the domain wall formation
and localization.  In the limit of large system sizes, the
corresponding time-dependent version of Eq.~(\ref{eq:mfa0}) which
governs the dynamics of the "coarse-grained" density $\rho$ reads
\begin{equation}
\label{eq:feuilleton}
\partial_t \rho + (1-2 \rho) \partial_x \rho = {\cal F}_A - {\cal F}_D \, .  
\end{equation}  
One can easily see that on hydrodynamic scales the source contribution
on the right hand side are negligible compared to the terms related to
the transport dynamics. On these scales, the local dynamics is
essentially described by mass conservation just as in the totally
asymmetric exclusion process. Neglecting the source contribution, one
can give an implicit analytic solution of Eq.~(\ref{eq:feuilleton}) by
standard methods of partial differential equations
\cite{zwillinger:book,zachmanoglou_thoe:book}. From such analysis, one
can infer the mechanism of the formation of density discontinuities
such as shocks on the hydrodynamic scale, which usually build up in
finite time.  An interesting feature of the domain wall is also its
slope $S_w$~\footnote{By a simple geometrical consideration, the slope
  is a measure for the inverse of the width.} as a function of the
system size $N$.  In Fig.~\ref{f:slope} we show the slope of the
domain wall as obtained from Monte-Carlo simulation for growing system
size $N$.  We find the scaling law $S_w \sim N^{\eta}$ with the
exponent $\eta = 0.50 \pm 0.05$.
This result is fully compatible with the scaling exponent $\eta=1/2$
computed for a pure totally asymmetric exclusion process. Note that
the mean-field approximation provides a wrong exponent $\eta_{MFA}=1$.
The softening of the slope compared to mean-field was recently
explained in Ref. ~\cite{evans_etal:03,popkov_etal:03} on the basis of domain wall
fluctuations~\cite{kolomeisky_schuetz_kolomeisky_straley:98}. In this
picture, the fluctuating domain wall performs a random walk just like
in the totally asymmetric exclusion process.  However, since on global
scale there is no mass conservation due to the Langmuir kinetics, the
current is space dependent and drives the domain wall to an
equilibrium position corresponding to a cusp in the current profile.
Such domain wall behavior can be rephrased in terms of a random walk
in the presence of a confining potential \cite{rakos_etal:03}. This
picture also suggests that the exponent for the slope of the domain
wall is exactly given by $\eta=1/2$.

It would be interesting to study how the case of equal rates can be
obtained by a limiting procedure of the case with slightly different
rates. The change of topology in the phase diagrams should be
contained in the analytic solution, however one suspects from
Eq.~(\ref{eq:gencoord}) that an essential singularity appears.
Furthermore, one would like to see if possible variants of our model
introduce new features similar to what has been done for a reference
dynamics, i.e. the totally asymmetric exclusion process
\cite{schuetz:review2}.

\acknowledgments This work was partially supported by the Deutsche
Forschungsgemeinschaft (DFG) under contract no. 850/4. A.P. was
supported by a Marie-Curie Fellowship no. HPMF-CT-2002-01529.


\end{document}